\documentclass[11pt]{article}

\usepackage[preprint]{acl}
\usepackage{booktabs, makecell}
\usepackage{enumitem}
\usepackage{amsmath, amssymb}
\usepackage{array}
\usepackage{booktabs}
\usepackage{multirow}
\usepackage{times}
\usepackage{latexsym}

\usepackage[T1]{fontenc}

\usepackage[utf8]{inputenc}

\usepackage{microtype}

\usepackage{inconsolata}

\usepackage{graphicx}
\usepackage[table]{xcolor}  
\usepackage{colortbl}       

\definecolor{em}{gray}{0.9}

\usepackage{colortbl}
\definecolor{bestbg}{RGB}
{255,235,235}   
\definecolor{secondbg}{RGB}{235,245,255} 
\usepackage{tabularx,booktabs}
\newcolumntype{L}[1]{>{\raggedright\arraybackslash}p{#1}}
\newcolumntype{Y}{>{\raggedright\arraybackslash}X}
\usepackage{xcolor}
\usepackage{soul}
\definecolor{lightpink}{HTML}{F7DADA}

\title{SAE-FiRE: Enhancing Earnings Surprise Predictions Through Sparse Autoencoder Feature Selection}

\author{Huopu Zhang\textsuperscript{1}, Yanguang Liu\textsuperscript{2}, Miao Zhang\textsuperscript{1}, Zirui He\textsuperscript{2}, \textbf{Mengnan Du\textsuperscript{2,*}}\\
\textsuperscript{1}Georgia Institute of Technology \,
\textsuperscript{2}New Jersey Institute of Technology\\
\small\texttt{\{hzhang931, jimmy.miao.zhang\}@gatech.edu}, \small\texttt{\{yanguang.liu, zh296, mengnan.du\}@njit.edu}\\[2pt]
\small\textsuperscript{*}Corresponding author
}

\begin{document}
\maketitle
\begin{abstract}
Predicting earnings surprises from financial documents, such as earnings conference calls, regulatory filings, and financial news, has become increasingly important in financial economics. However, these financial documents present significant analytical challenges, typically containing over 5,000 words with substantial redundancy and industry-specific terminology that creates obstacles for language models. In this work, we propose the \textbf{SAE-FiRE} (Sparse Autoencoder for Financial Representation Enhancement) framework to address these limitations by extracting key information while eliminating redundancy. SAE-FiRE employs Sparse Autoencoders (SAEs) to decompose dense neural representations from large language models into interpretable sparse components, then applies statistical feature selection methods, including ANOVA F-tests and tree-based importance scoring, to identify the top-k most discriminative dimensions for classification. By systematically filtering out noise that might otherwise lead to overfitting, we enable more robust and generalizable predictions. Experimental results across three financial datasets demonstrate that SAE-FiRE significantly outperforms baseline approaches.
\end{abstract}


\section{Introduction}

Financial documents such as earnings conference calls, regulatory filings, and financial news provide investors with retrospective assessments of performance and forward-looking guidance. Executives and other intermediaries often possess material nonpublic information, and their language, explicit or implicit, can reveal valuable insights about firms’ prospects. These sources form a rich set of unstructured data for forecasting earnings surprises and market reactions~\citep{brown2004conference,bushee2018linguistic}.


Despite their informational value, financial documents present substantial challenges for automated analysis. The primary obstacles stem from their considerable length (often exceeding 5,000 words) and the presence of significant redundancy, boilerplate language, and industry-specific jargon~\citep{loughran2011liability}. These characteristics create noise that can obscure genuinely predictive signals. These deep neural network models can easily overfit to these noises, focusing on spurious correlations rather than meaningful financial indicators. This overfitting tendency leads to poor generalization ability of the prediction models, significantly limiting their practical utility in real-world  scenarios.

Sparse Autoencoders (SAEs) have recently gained increasing attention in the large language model (LLM) and explainability community as tools for understanding the internal representations of LLMs~\citep{lieberum2024gemma,he2024llama}. Originally developed to enhance model interpretability, SAEs decompose dense neural representations into interpretable sparse components, effectively disentangling complex features. By identifying and isolating essential patterns within neural activations, SAEs provide insights into how LLMs process and represent information. This capability makes them particularly suited for addressing the challenges of financial documents analysis, where distinguishing between predictive signals and redundant noise is crucial. 

In this work, we propose a novel methodology for robust earnings surprise prediction via \textbf{SAE-FiRE} (Sparse Autoencoder for Financial Representation Enhancement). Our approach addresses the limitations of existing methods by: (1) extracting key information while eliminating redundancy from lengthy documents, and (2) employing SAEs to efficiently process and identify patterns within LLM representations. 
By systematically filtering out noise that might otherwise lead to overfitting, we enable more robust and generalizable predictions through a novel feature selection approach that employs two statistical methods: ANOVA F-tests, and tree-based method to identify only the most discriminative SAE dimensions. Experimental results across three financial datasets (earnings call transcripts, 10Q reports, and financial news) demonstrate that SAE-FiRE significantly outperforms baseline approaches.
Our contribution in this work can be summarized as follows:
\begin{itemize}[leftmargin=*]\setlength\itemsep{-0.3em}
\item We propose the SAE-FiRE framework for earnings surprise prediction that effectively suppresses feature noise in financial documents.
\item We develop a systematic feature selection methodology using two statistical criteria (F-test, and tree-based methods) to identify the most discriminative SAE dimensions.
\item Experiments indicate that SAE-FiRE significantly outperforms baseline models, demonstrating improved robustness and generalizability in earning surprises predictions.
\end{itemize}

\begin{figure*}[ht]
  \centering
  \includegraphics[width=1\textwidth]{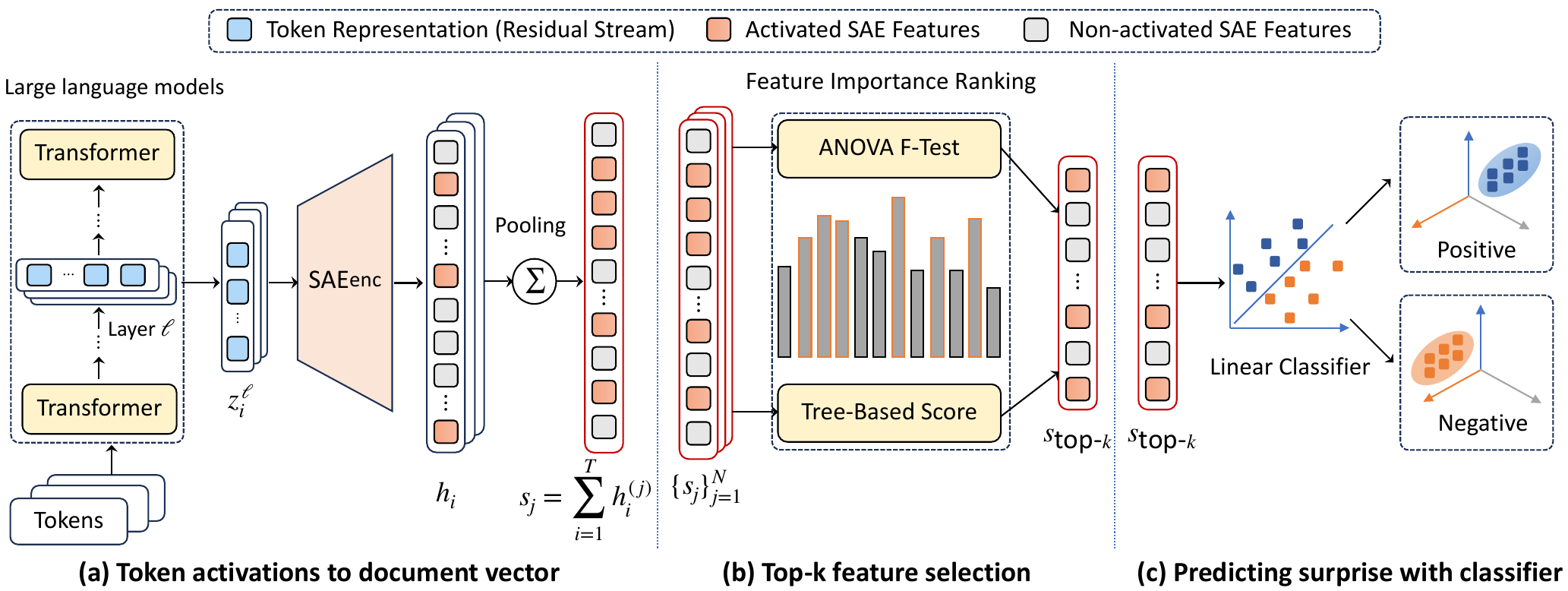}
  \caption{Diagram of our SAE‐based earnings‐surprise prediction pipeline. (a) extracting and pooling token-level SAE activations into a document representation, (b) selecting the most predictive features using statistical ranking methods, and (c) training a linear classifier on the filtered feature set.}
  \label{fig:method-arch}
\end{figure*}

\section{Preliminaries}

\subsection{LLM Hidden Representations}
We focus on decoder-only LLMs (GPT-like models), which process information through sequential layers of multi-head attention blocks (MHA) and feed-forward networks (FFNs/MLPs). In our approach, we utilize pre-trained LLMs as frozen feature extractors without fine-tuning their parameters. 
Each layer $\ell \in L$ in these models operates on a residual stream, where information flows through the network while maintaining direct connections to previous states. The computation at each layer follows a consistent pattern: the MHA mechanism first processes the input and adds its output back to the residual stream, which is then transformed by an MLP to generate the layer's final output:
\begin{align}
\small
\boldsymbol{z}_i^{\ell+1} = \boldsymbol{z}_i^{\ell} + \text{Att}^{\ell} (\boldsymbol{z}_i^{\ell}) + \text{MLP}^{\ell} \left( \boldsymbol{z}_i^{\ell} + \text{Att}^{\ell} (\boldsymbol{z}_i^{\ell}) \right),
\end{align}
where $\boldsymbol{z}_i^{\ell}$ represents the representation of the $i$-th token at layer $\ell$. 
For our analysis, we extract hidden representations from the residual stream at each layer's output. 


\subsection{Sparse Autoencoders (SAEs)}
After obtaining hidden representations from the LLM, we employ SAEs to disentangle these dense vectors into interpretable concepts through dictionary learning. SAEs transform the original representation into a higher-dimensional but highly sparse activation space, enabling more effective isolation of meaningful patterns.

For a hidden representation $\boldsymbol{z}^{\ell} \in \mathbb{R}^d$ (simplified as $\boldsymbol{z}$ for clarity) from the LLM's residual stream, an SAE performs encoding followed by decoding to reconstruct the original input: $\boldsymbol{z} = \text{SAE}(\boldsymbol{z}) + \epsilon$, where $\epsilon$ represents the reconstruction error. In our framework, we implement layer-wise SAEs consisting of an encoder matrix $\boldsymbol{W}_\text{enc} \in \mathbb{R}^{d \times m}$, a decoder matrix $\boldsymbol{W}_\text{dec} \in \mathbb{R}^{m \times d}$, and a non-linear activation function \citep{he2024llama}. The forward pass through the SAE can be defined as:
\begin{align}
&h(\boldsymbol{z})=\sigma\left(\boldsymbol{z} \boldsymbol{W}_{\mathrm{enc}}\right.+\left.\boldsymbol{b}_{\mathrm{enc}}\right), \label{eq:1} \\
&\operatorname{SAE}(\boldsymbol{z})=h(\boldsymbol{z}) \boldsymbol{W}_{\mathrm{dec}}\ +\ \boldsymbol{b}_{\mathrm{dec}}, \label{eq:2}
\end{align}
where $\boldsymbol{b}_\text{enc} \in \mathbb{R}^m$ and $\boldsymbol{b}_\text{dec} \in \mathbb{R}^d$ are bias terms, and $h(\boldsymbol{z})$ represents the sparse latent activations in the expanded dimension $m \gg d$.

Different SAE implementations utilize different sparsity-inducing activation functions $\sigma$, such as TopK-ReLU, JumpReLU \citep{rajamanoharan2024jumpingaheadimprovingreconstruction}, etc. 
In the context of financial textual disclosures, SAE provides us an effective tool to separate predictive financial signals from redundant noise that might otherwise confuse the model.

\section{Proposed Method}

We propose SAE-FiRE, which leverages the inherent capacity of SAEs to identify and isolate essential patterns while filtering out noise.  Our SAE-FiRE  systematically removes redundant noise that might otherwise lead to overfitting, ultimately producing more robust and generalizable earnings surprise predictions. SAE-FiRE contains the following three steps (see Figure~\ref{fig:method-arch}).

\subsection{Obtaining Representations} 
For every financial document (truncated to \(T=20{,}000\) tokens), we extract each token's residual-stream activations from layers instrumented with pretrained SAEs and the SAELens tool~\citep{bloom2024saetrainingcodebase}.  We denote the $m$-dimensional SAE feature vector at token position $i$ as $\mathbf{h}_i \in \mathbb{R}^m$, for $i=1,2,\dots,T$.
To aggregate the token‐wise representations into a single, fixed‐length embedding for the entire document, we apply a simple sum‐pooling operator:
\begin{equation}
\mathbf{s}
\;=\;
\sum_{i=1}^{T} \mathbf{h}_i
\quad\in\;\mathbb{R}^m.
\label{eq:sum_pooling}
\end{equation}
Here, \(\mathbf{s}\) serves as the global SAE signature of the document, aggregating activation strength across all token positions without any learned parameters. This summation‐based pooling not only enforces a consistent input size for downstream classifiers but also emphasizes the cumulative contribution of each latent feature over the full document.

\subsection{Feature Selection via SAE}  
After sum‐pooling, each financial document is represented by a \(m\)–dimensional vector \(\mathbf{s}\). Each financial document is paired with its Standardized Unexpected Earnings (SUE) score based on the earnings per share (EPS) reported in the immediately following quarter from the earnings surprise database. We labeled the samples with \(\text{SUE}\ge0.5\) as `positive', denoting positive earnings surprise, and \(\text{SUE}\le-0.5\) as `negative', denoting negative earnings surprise, discarding those in \((-0.5,0.5)\) as non-surprise.  To quantify the ability of each SAE feature to separate two classes, we compute two univariate scores:
\begin{itemize}[leftmargin=*]\setlength\itemsep{-0.3em}
  \item \(F\)–test (ANOVA).  
    Viewing the two groups as levels of a factor, we compute the \(F\)-statistic, which compares between-group variance to within-group variance:
    \begin{equation}
    \small
      F_j \;=\; \frac{\bigl(\bar{s}_{j,+}-\bar{s}_j\bigr)^2 + \bigl(\bar{s}_{j,-}-\bar{s}_j\bigr)^2}{\tfrac1{n_+ + n_- -2}\sum_{i}(s_{j}^{(i)}-\bar{s}_{\text{group}})^2},  
    \end{equation}
    where \(\bar{s}_j\) is the overall mean.  Higher \(F_j\) signals greater dispersion between classes, though it can be sensitive to outliers.


    \item Tree‐Based Method.
    We train an ensemble of decision trees using gradient boosting to predict the binary surprise label \(Y\) from the full activation vector \(\mathbf{s}\). For each feature \(j\), we record its average importance score, such as mean decrease in impurity or permutation importance, across all trees. Features with higher importance contribute more to reducing prediction error and may capture complex, non‐linear interactions.
\end{itemize}

Each method produces a score per feature. We then rank the \(m\) features accordingly and retain the top-\(k\) for downstream classification to reduce noise and facilitate efficient classifier training. By filtering out redundant and spurious activations, our approach yields more robust and generalizable feature sets. We experimented with feature selection using each of these two methods to assess their relative effectiveness.

\subsection{Fitting A Classifier}  
Using the top-\(k\) SAE features ranked by our selection criteria, we train logistic regression classifiers to predict positive versus negative earnings surprises.  For each document, the pooled SAE vector \(\mathbf{s}\) is reduced to its \(k\) most discriminative dimensions, and these \(k\) features are used as input to a logistic regression model with an \(\ell_2\)-regularized loss.
The objective function to minimize is:
\begin{equation}
\small
\mathcal{L}(\mathbf{w}, b) = -\frac{1}{N} \sum_{i=1}^{N} [y_i \log(\hat{y}_i) + (1-y_i) \log(1 - \hat{y}_i)] + \frac{\lambda}{2} \|\mathbf{w}\|_2^2
\label{eq:logistic_loss}
\end{equation}
where $y_i \in \{0, 1\}$ is its true label, $\hat{y}_i = \sigma(\mathbf{w}^T s'_i + b)$ is the predicted probability, and $\lambda$ is the  parameter controlling the strength of the \(\ell_2\) penalty.
All classifiers are trained and evaluated using a chronological train/validation/test split based on cutoff dates, and hyperparameters are selected via grid search on the validation set. 











\begin{table*}[t]
\centering
{\fontsize{8}{9}\selectfont          
\setlength{\tabcolsep}{3pt}          
\renewcommand{\arraystretch}{0.82}   

\begin{tabular*}{\textwidth}{@{\extracolsep{\fill}} l ccc ccc ccc @{}}
\toprule
\textbf{Statistic} &
\multicolumn{3}{c}{\textbf{Conference Call}} &
\multicolumn{3}{c}{\textbf{10\,-Q Report}} &
\multicolumn{3}{c}{\makecell{\textbf{FNSPID}\\\textbf{Nasdaq News}}} \\
\cmidrule(lr){2-4} \cmidrule(lr){5-7} \cmidrule(lr){8-10}
& \rule{0pt}{1.05em} Train & Val & Test 
& \rule{0pt}{1.05em} Train & Val & Test 
& \rule{0pt}{1.05em} Train & Val & Test \\

\midrule
Start Date          & Jan 2012 & Jan 2014 & Jul 2014 & Jan 2012 & Jan 2014 & Jul 2014 & Jan 2019 & Jan 2021 & Jul 2021 \\
End Date            & Dec 2013 & Jun 2014 & Dec 2014 & Dec 2013 & Jun 2014 & Dec 2014 & Dec 2020 & Jun 2021 & Dec 2021 \\
\# Documents        & 6147 & 1788 & 1389 & 3135 & 561 & 1107 & 2276 & 652 & 731 \\
Avg \# of Words     & 8374 & 8269 & 8685 & 8240 & 7320 & 8447 & 4326 & 4310 & 4509 \\
Median \# of Words  & 8097 & 7861 & 8365 & 7151 & 6669 & 7468 & 2219 & 2229 & 2604 \\
Max \# of Words     & 58448 & 66240 & 68537 & 75236 & 41707 & 52036 & 44746 & 55853 & 43312 \\
Avg \# of Sentences & 440 & 439 & 464 & 266 & 237 & 269 & 121 & 134 & 145 \\
Avg \# of Words per Sentence & 19 & 19 & 19 & 31 & 31 & 31 & 36 & 32 & 31 \\
\bottomrule
\end{tabular*}
}
\caption{Summary statistics of the three datasets used in our experiments.}
\label{tab:summary_stats}
\end{table*}

\section{Experiments}
In this section, we evaluate the SAE-FiRE to answer the following research questions (RQs):
\begin{itemize}[leftmargin=*]\setlength\itemsep{-0.3em}
    \item \textbf{RQ1:} How does the performance of SAE-FiRE compare to baseline models?  (Section 4.2)
    \item \textbf{RQ2:} How does feature importance ranking methods and the number of selected features influence the performance? (Section 4.3)
    \item \textbf{RQ3:} How does language model size and layer influence the performance? (Section 4.4 and 4.5)
    \item \textbf{RQ4:} Does our SAE-FiRE framework have the benefit of interpretability ? (Section 4.6)
\end{itemize}

\subsection{Experimental Settings}

\noindent\textbf{Earnings Surprise Datasets.}
We employ three main datasets in our analysis: (i) Conference Call, (ii) 10\,-Q Report, and (iii) FNSPID Nasdaq News. These datasets capture complementary dimensions of firm information environments, including managerial communication, formal regulatory disclosures, and media narratives, providing a rich empirical foundation for our experiments. Summary statistics are presented in Table~\ref{tab:summary_stats}. Please refer to details of the datasets in Section~\ref{appendix:sec:datasets} in Appendix.

\vspace{3pt}
\noindent\textbf{Supervised Learning Task.}
We frame the prediction of the direction of the next earnings surprise (\( ES \)) as a supervised learning task using only the textual content of the most recent financial documents. Following \citet{latane1979standardized}, we measure \( ES \) using the Standardized Unexpected Earnings (SUE), defined as the difference between reported EPS and the analyst consensus estimate, scaled by the standard deviation of analyst forecasts. 
The consensus estimate is computed as the mean of the latest valid analyst forecasts issued within one month after the earnings call, allowing analysts to update their expectations based on the call content and recent financial disclosures. This setup provides a forward-looking measure of market expectations and presents a more realistic yet challenging prediction task. The average time horizon between the input document and the target earnings event is approximately three months, further underscoring the difficulty of the task.
\begin{equation}
\small
ES = \frac{RepEPS - \text{Avg}(EstEPS)}{\text{Std}(EstEPS)}
\end{equation}
\begin{equation}
\small
y =
\begin{cases}
0, & ES \leq -\delta \\
1, & ES \geq \delta
\end{cases}
\end{equation}
We transform the continuous earnings surprise (\( ES \)) into a binary classification task by assigning a label of \( +1 \) for \( ES > \delta \) and \( 0 \) for \( ES < -\delta \), where \( \delta = 0.50 \). This choice follows previous studies on standardized unexpected earnings \citep{eli_bartov_patterns_1992} and on price momentum \citep{luo_retail_2022} , where earnings surprise are considered large when \(|SUE| \ge 0.5\).This threshold balances sample size and event significance. Documents with immaterial surprises (i.e., \( ES \in [-0.50, 0.50] \)) are excluded, as these near-zero values often elicit weak market responses and may reflect earnings management. We evaluate performance using accuracy and train models with binary cross-entropy loss.
Although \( ES \) is a continuous variable, market reactions are typically binary, responding more to the direction of the surprise than to its magnitude. Our focus is therefore on material surprises that are more likely to influence investor behavior and pricing.

\vspace{3pt}
\noindent\textbf{Models and SAE.} 
We consider two LLM families, Gemma and Llama. For Gemma, we use two scales of the Gemma2 model, Gemma2-2B and Gemma2-9B. For Llama, we consider Llama 3.1-8B. We use SAELens~\citep{bloom2024saetrainingcodebase} to extract residual-stream activations at selected layers and to apply the corresponding pretrained SAEs for each model, obtaining activations in the SAE latent space.

\vspace{3pt}
\noindent\textbf{Feature Selection.}  
To reduce noise and highlight the most discriminative SAE dimensions, we rank each feature by two univariate criteria:  
(1) $F$–test (ANOVA),  
(2) Tree-based Methods.  
For each criterion, features are scored and sorted, and we retain the top $k\in\{500, 1000, 1500, 2000, 2500\}$ dimensions with Gemma 2-2B 16K SAE (where 16K indicates the SAE has 16,384 features, i.e., $m$=16K), and the top $k\in\{3000, 3500, 4000, 4500, 5000, 5500, 6000\}$ dimensions with Gemma 2-9B 131K SAE and Llama 3.1-8B 131k SAE. For each SAE, we evaluate all combinations of feature selection method and $k$ value on the validation set, and select the best-performing combination  as our final model configuration for test set evaluation.

\vspace{3pt}
\noindent\textbf{Classifier Training.}  
On the selected top-$k$ features, we train $\ell_2$‐regularized logistic regression models to predict the binary surprise label.  Regularization strength is tuned on the validation set via grid search, and final performance is measured on the test set, reporting accuracy, F1 score, and AUC.

\vspace{3pt}
\noindent\textbf{\textcolor{black}{Baselines.}}  
We compare with 8 baselines:

\noindent
\underline{\textcolor{black}{Zero-shot prompting on LLM.}} 
    We apply the Gemma 3‑12B-IT model in a pure zero‐shot setting. For each input, we prepend a task‑specific instruction and decode greedily (temperature=0). No demonstrations or parameter updates are used.

\noindent
\underline{\textcolor{black}{Few-shot prompting on LLM.}}
    Because Gemma 2’s context window cannot hold multiple examples plus the query, we instead use Gemma 3‑12B-IT with \(k=4\) in‑context demonstrations sampled from the training set, followed by the test input. Decoding remains greedy.

\noindent
\underline{\textcolor{black}{Longformer.}}
    We fine‐tune the Longformer‐base‐4096 model \citep{beltagy2020longformer}. Please refer to Section~\ref{appendix-sec-baselines} in Appendix for more details.

\noindent
\underline{\textcolor{black}{Hierarchical FinBERT.}}
    We implement the Hierarchical FinBERT model from \citep{koval2023forecasting} for long-document  classification. Please refer to Section~\ref{appendix-sec-baselines} in Appendix for more details.

\noindent
\underline{SAE without feature selection.}  
    Using the same pooled SAE vectors \(\mathbf{s}\) (Gemma 2-2B at \(d=16\text{K}\); Gemma 2-9B at \(d=131\mathrm{K}\), we omit the top-\(k\) selection step and train:
    \vspace{-5pt}
    \begin{itemize}[leftmargin=*]\setlength\itemsep{-0.3em}
      \item \emph{XGBoost.}  max number of boosting rounds 1000, learning rate 0.1, max depth 6, subsample 0.8, with early stopping on the validation AUC (patience = 20 rounds).
      \item \emph{MLP.}  Three hidden layers of sizes [256, 256, 256], ReLU activations, dropout 0.5, trained with AdamW (learning rate \(1\times10^{-6}\)), batch size 32, for up to 200 epochs with early stopping.
    \end{itemize}

\noindent
  \underline{Last‐hidden‐state probes.}  
    We extract the final‐token hidden representation \(\mathbf{h}_{\mathrm{last}}\in\mathbb{R}^{d_{\mathrm{hid}}}\) from the last layer of each Gemma 2 model (2B or 9B).  On these vectors we train:
    \begin{itemize}[leftmargin=*]\setlength\itemsep{-0.3em}
      \item \emph{Logistic Regression.}  \(\ell_2\)-regularization with inverse penalty \(C\in\{0.0001, 0.001,0.01,0.1,1\}\), selected by validation AUC.
      \item \emph{MLP.}  One hidden layer of size 256, ReLU activations, dropout 0.2, optimized with AdamW (learning rate \(1\times10^{-6}\)), 
      up to 200 epochs with early stopping.
    \end{itemize}

\begin{table*}[tp]
\centering
\setlength{\tabcolsep}{3pt}
\caption{Performance comparison of SAE-FiRE framework against baseline methods across three key metrics. For SAE-FiRE-Gemma 2-2B, results are reported with $k=1500$ selected features; for SAE-FiRE-Gemma 2-9B and SAE-FiRE-Llama 3.1-8B, results use $k=4500$ selected features. \colorbox{bestbg}{\textbf{Red}}: the best,  \colorbox{secondbg}{\textbf{Blue}}: the second best.}
\label{tab:sae_fire_comparison_fullwidth}
\resizebox{1.0\textwidth}{!}{%
\begin{tabular}{@{}l 
  c c c 
  c c c 
  c c c@{}}
\toprule
\multirow{2}{*}{Method} 
  & \multicolumn{3}{c}{Conference Call} 
  & \multicolumn{3}{c}{10Q Report } 
  & \multicolumn{3}{c}{FNSPID Nasdaq News} \\
\cmidrule(lr){2-4} \cmidrule(lr){5-7} \cmidrule(lr){8-10}
& {Accuracy}$\uparrow$ & {F1-Score}$\uparrow$ & {ROC AUC}$\uparrow$ 
  & {Accuracy}$\uparrow$ & {F1-Score}$\uparrow$ & {ROC AUC}$\uparrow$ 
  & {Accuracy}$\uparrow$ & {F1-Score}$\uparrow$ & {ROC AUC}$\uparrow$ \\
\midrule
Zero-shot Prompting & 0.650 & 0.676 & 0.546 & 0.596 & 0.628 & 0.531 & 0.647 & 0.669 & 0.558 \\
Few-shot Prompting & 0.694 & 0.695 & 0.569 & 0.611 & 0.642 & 0.549 & 0.685 & 0.689 & 0.567 \\
Longformer & 0.744 & 0.718 & 0.590 & 0.721 & 0.704 & 0.587 & 0.758 & 0.743 & 0.609 \\
Hierarchical FinBERT & 0.772 & 0.721 & 0.613 & 0.735 & 0.712 & 0.602 & 0.771 & 0.754 & 0.623 \\
Gemma 2 Last Hidden + LR & 0.761 & 0.737 & 0.628 & 0.742 & 0.719 & 0.611 & 0.784 & 0.766 & 0.636 \\
Gemma 2 Last Hidden + MLP & 0.770 & 0.731 & 0.634 & 0.748 & 0.718 & 0.616 & 0.791 & 0.764 & 0.624 \\
SAE 16k Features + XGBoost & 0.781 & 0.715 & 0.620 & 0.735 & 0.713 & 0.636 & 0.798 & 0.758 & 0.613 \\
SAE 16k Features + MLP & 0.773 & 0.737 & 0.642 & 0.747 & 0.726 & 0.621 & 0.805 & 0.763 & 0.628 \\
\midrule
{\textbf{SAE-FiRE-Gemma 2-2B (Ours)}} & 0.793 & 0.743 & \textbf{0.657} & \textbf{0.757} & \cellcolor{secondbg}\textbf{0.741} & \textbf{0.668} & \textbf{0.815} & \textbf{0.772} & \textbf{0.674} \\
{\textbf{SAE-FiRE-Gemma 2-9B (Ours)}} & \cellcolor{bestbg}\textbf{0.801} & \cellcolor{secondbg}\textbf{0.757} & \cellcolor{secondbg}\textbf{0.668} & \cellcolor{bestbg}\textbf{0.771} & \cellcolor{bestbg}\textbf{0.758} & \cellcolor{secondbg}\textbf{0.680} & \cellcolor{bestbg}\textbf{0.827} & \cellcolor{bestbg}\textbf{0.798} & \cellcolor{bestbg}\textbf{0.703} \\
{\textbf{SAE-FiRE-Llama 3.1-8B (Ours)}} & \cellcolor{secondbg}\textbf{0.798} & \cellcolor{bestbg}\textbf{0.759} & \cellcolor{bestbg}\textbf{0.673} & \cellcolor{secondbg}\textbf{0.769} & \textbf{0.732} & \cellcolor{bestbg}\textbf{0.689} & \cellcolor{secondbg}\textbf{0.822} & \cellcolor{secondbg}\textbf{0.788} & \cellcolor{secondbg}\textbf{0.700} \\
\bottomrule
\end{tabular}%
}
\end{table*}

\vspace{3pt}
\noindent\textbf{Implementation Details.}  
SAE extraction leverages the SAELens plugin within the transformer inference pipeline. Specifically, for Gemma 2-2B we extract residual‐stream activations at layer 12 (middle layer) using a pretrained SAE of width \(m=16\mathrm{K}\); for Gemma 2-9B, we extract at layer 20 (middle layer) with SAE widths \(m=131\mathrm{K}\); for Llama 3.1-8B, we extract at layer 20 (middle layer) with SAE widths \(m=131\mathrm{K}\). Activation pooling and feature selection routines use NumPy, scikit‐learn (1.6.1), and XGBoost (3.0.0). Logistic regression and XGBoost baselines are trained with scikit‐learn and XGBoost, while MLPs employ PyTorch with three hidden layers of size 256 and dropout of 0.5 and ReLU activations.
We ran all extraction and training on a cluster of three NVIDIA H100 GPUs under CUDA 12.6 and Ubuntu 20.04.  Batch size for activation extraction was 1 document (capped to 20 000 tokens), and all random seeds were fixed to 42 for reproducibility.

\subsection{Comparison with Baseline Models}
We compare SAE-FiRE with baselines across three datasets: Conference Call transcripts, 10Q Reports, and FNSPID Nasdaq News. Results are reported in Table~\ref{tab:sae_fire_comparison_fullwidth}, yielding the following key insights:

\vspace{3pt}
\noindent\textbf{Overall Performance Rankings.} 
Across all three datasets and metrics, SAE-FiRE variants consistently occupy the top three positions. On Conference Calls, the three SAE-FiRE models achieve the highest F1-scores (0.759, 0.757, 0.743) and AUC values (0.673, 0.668, 0.657). On 10Q Reports, they similarly dominate with F1-scores of 0.758, 0.741, and 0.732, and AUC values of 0.689, 0.680, and 0.668. On FNSPID News, they reach F1-scores of 0.798, 0.788, and 0.772, with AUC values of 0.703, 0.700, and 0.674. This consistent superiority demonstrates the robustness and generalizability of our approach across different types of financial text, from structured regulatory filings to conversational earnings calls to news articles.

\vspace{3pt}
\noindent\textbf{Comparison to In-Context Learning.}  
In-context learning approaches consistently underperform across all datasets. On Conference Calls, zero-shot prompting achieves 0.676 F1 and 0.546 AUC, while few-shot prompting (with 4 examples) improves to 0.695 F1 and 0.569 AUC. Similar patterns emerge on 10Q Reports. In all cases, the reliance on limited demonstrations prevents these models from effectively capturing the complex discourse structures and nuanced contextual cues in lengthy financial documents, causing performance to lag substantially behind long-document models and SAE-FiRE variants.

\vspace{3pt}
\noindent\textbf{Comparison to Long-Document Models.}  
Longformer and Hierarchical FinBERT represent strong baselines, but SAE-FiRE consistently outperforms both. On Conference Calls, Longformer achieves 0.718 F1 (0.590 AUC) and Hierarchical FinBERT reaches 0.721 F1 (0.613 AUC), compared to SAE-FiRE's 0.743-0.759 F1 (0.657-0.673 AUC). The pattern holds on both 10Q Reports and FNSPID News. While long-document models extend context windows, they uniformly process all segments and often dilute key signals with irrelevant content. By contrast, SAE-FiRE pools activations over entire transcripts and selectively retains only the top-k most discriminative features, yielding more focused representations and superior performance.

\begin{figure*}[htb]
  \centering
  \includegraphics[width=1\textwidth]{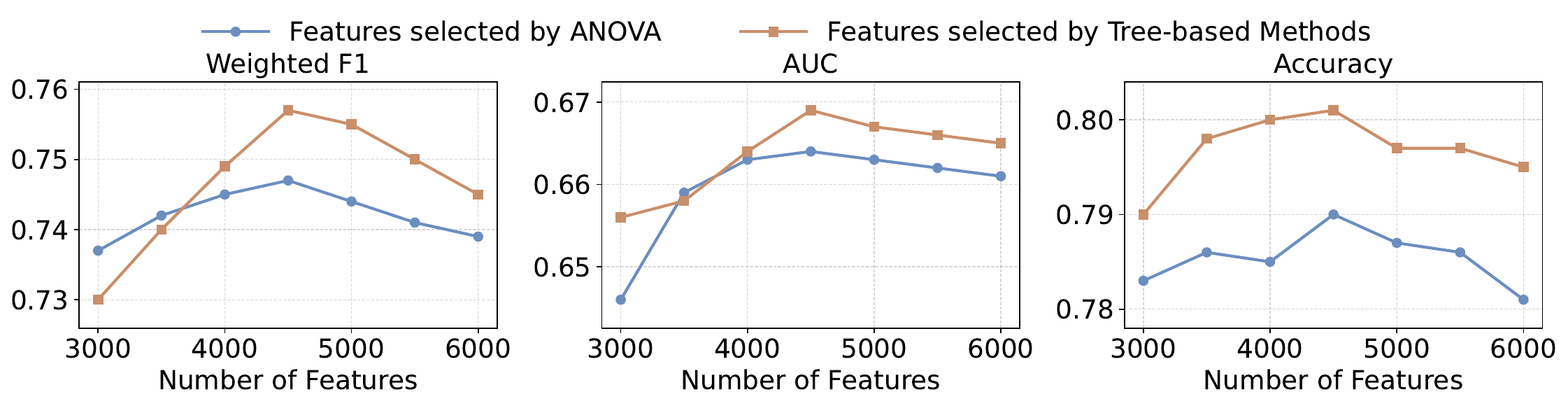}
  \caption{
    Weighted F1, AUC, and Accuracy at different numbers of selected features. (Gemmma2-9B 131K)
  }
  \label{fig:num-feature-comparison1}
  \label{fig:feature-selection}
\end{figure*}

\vspace{3pt}
\noindent\textbf{Comparison to Hidden State Probes.}
Probing Gemma 2's last hidden states directly yields weaker results than SAE-FiRE across all datasets and classifiers. On Conference Calls, logistic regression achieves 0.737 F1 (0.628 AUC) and MLP reaches 0.731 F1 (0.634 AUC). In comparison, SAE-FiRE consistently achieves 0.732-0.759 F1 on Conference Calls. This demonstrates that sum-pooled SAE activations capture cumulative, global patterns across full documents more effectively than final-token embeddings, which miss broader contextual cues. Additionally, our univariate feature selection prunes spurious dimensions, sharpening class boundaries and improving generalization.

\vspace{3pt}

\noindent\textbf{Benefit of Selecting Top SAE Features.}
Training directly on full SAE activations without feature selection yields suboptimal results across all datasets. For example on Conference Calls, using all 16K features with MLP achieves 0.737 F1 (0.642 AUC) and XGBoost reaches 0.715 F1 (0.620 AUC). By contrast, SAE-FiRE with selective top-k feature retention achieves 0.743 F1 (0.657 AUC) on Conference Calls using the Gemma 2-2B variant. This confirms that filtering to the most discriminative features reduces overfitting to document-specific noise while enabling faster training and lower memory requirements.

\vspace{3pt}
\subsection{Influence of Feature Selection}
In this section, we investigate the influence of feature importance ranking methods on the Conference Call dataset. We report the results in Figure~\ref{fig:feature-selection}.

\vspace{3pt}
\noindent\textbf{Comparison of Feature Selection Methods.}
We compare the model performance using two different feature selection methods: ANOVA F-tests, and tree-based method. The result shows that models trained with features selected using the tree-based approach outperforms models using features selected using ANOVA for both the model based on 16K SAE from Gemma 2-2B and the model based on 131K SAE from Gemma 2-9B. For example, for SAE-FiRE with Gemma 2-9B 131K SAE, the weighted F1 is 0.757, the AUC is 0.668, and the accuracy is 0.801 when using tree-based method, compared to the weighted F1 of 0.748, AUC of 0.664, and accuracy of 0.790 when using ANOVA. For SAE-FiRE with Gemma 2-2B 16K SAE, the weighted F1 is 0.743, the AUC is 0.657, and the accuracy is 0.793 when using tree-based method, compared to the weighted F1 of 0.743, AUC of 0.646, and accuracy of 0.788 when using ANOVA. 

\vspace{3pt}
\noindent\textbf{Comparison of Number of Features.}
For SAE-FiRE with Gemma 2-2B 16K, we test the number of top features $k\in\{500, 1000, 1500, 2000, 2500\}$, and we observe a peak of performance at around $k = 1500$. For Gemma 2-9B 131K , we test $k\in\{3000, 3500, 4000, 4500, 5000, 5500, 6000\}$, and we observe a peak of performance at around $k = 4500$. These findings suggest that \emph{there is an optimal feature subset size}. When the number of features is smaller, the model cannot access all of the discriminative signals by omitting useful predictors, whereas when too many less important features are introduced, noises increase, which has a negative effect on classification.

\begin{table}[t]
  \centering
  \small
  \caption{Comparison of Accuracy, AUC, and Weighted Average F1 Between Gemma 2-2B and Gemma 2-9B}
  \label{tab:all_metrics_comparison}
  \scalebox{0.9}{\begin{tabular}{lccc}
    \toprule
    \textbf{Model} 
      & \textbf{Accuracy} 
      & \textbf{AUC} 
      & \textbf{F1} \\
    \midrule
    Last Hidden State + LR  
      & 0.761 
      & 0.628 
      & 0.737  \\
    Last Hidden State + MLP 
      & 0.770 
      & 0.634 
      & 0.731 \\
    \textbf{SAE-FiRE (Gemma2-2B)} 
      & \textbf{0.793} 
      & \textbf{0.657} 
      & \textbf{0.743}  \\
    \textbf{SAE-FiRE (Gemma2-9B)} 
      & \textbf{0.801} 
      & \textbf{0.668} 
      & \textbf{0.757} \\
    \bottomrule
  \end{tabular}}
  \label{tab:model-size}
\end{table}

\subsection{Influence of Model Size}
In this section, we study the influence of Gemma 2 model size on SAE feature quality using the Conference Call dataset, and the results are given in Table~\ref{tab:model-size}.
The SAE-FiRE framework using Gemma 2-9B 131K outperforms the same framework with Gemma 2-2B 16K across all key metrics, with the model using 9B 131K SAE bringing accuracy from 0.793 to 0.801, AUC from 0.657 to 0.668, and weighted average F1 from 0.743 to 0.757. The best performance is achieved when tree-based feature importance ranking methods are used. When Gemma 2-9B 131K is used, the best performance is achieved with top 4500 features and when Gemma 2-2B 16K is used, the best performance is achieved with top 1500 features. The results suggest that \emph{the larger model's richer representations enable the framework to pick out more predictive signals}. This confirms the value of larger models for capturing more subtle linguistic patterns to improve the prediction performance.


\subsection{Influence of Model Layers}
To assess how the depth of transformer layers affects the quality of SAE-derived representations for earnings surprise prediction, we perform experiments on Gemma2-2B model using the Conference Call dataset by extracting SAE activations from three distinct layers: an early layer (Layer 5), a middle layer (Layer 12), and a late layer (Layer 20). The results are shown in Table \ref{tab:sae_layers}.

\sethlcolor{lightpink}

\begin{table*}[t]
  \centering
  \begingroup
  \setlength{\tabcolsep}{9pt}
  \renewcommand{\arraystretch}{1.25}
  {\fontsize{10pt}{12pt}\selectfont
  \begin{tabularx}{\linewidth}{@{} L{2.6cm} L{1.6cm} Y @{}}
    \toprule
    \textbf{Model} & \textbf{Feature} & \textbf{Interpretation (Neuronpedia)} \\
    \midrule
    \multirow{5}{*}{Gemma 2-2B}
      & 4334 & legal terms and phrases related to contracts and warranties \\
      & 3820 & financial terms and references to wealth or monetary values \\
      & 1608 & phrases related to affirmation, recommendations, and positive evaluations \\
      & 5096 & references to urban environments and their characteristics \\
      & 6751 & terms related to political processes and events \\
    \addlinespace[2pt]
    \cmidrule(lr){1-3}
    \multicolumn{3}{@{}l}{\textbf{Example sentence}} \\
    \cmidrule(lr){1-3}
    \multicolumn{3}{@{}p{\dimexpr\linewidth-2\tabcolsep\relax}@{}}{
      It was a perfect \hl{example} of how a legally binding federal investment \hl{contract} could boost corporate profits and \hl{urban redevelopment},
      with government legislation ensuring regulatory compliance and sustained capital \hl{growth} across major \hl{metropolitan} \hl{markets}.
    }\\
    \bottomrule
  \end{tabularx}}
  \endgroup
  \caption{Top SAE activation features and their semantic interpretations for the \textbf{Gemma 2-2B model (Layer 12, 16K-width activations)}. Highlighted spans in the example sentence indicate words or phrases that strongly activate the corresponding SAE features.}
  \label{tab:gemma2_2b_features}
\end{table*}

\begin{table}[t]
  \centering
  \small
  \caption{Comparison Between SAE Activations from different Layers of Gemma 2-2B 16K SAE.}
  \label{tab:sae_layers}
  \setlength{\tabcolsep}{11.5pt}
  \begin{tabular}{lccc}
    \toprule
    \textbf{Layer} & \textbf{Accuracy} & \textbf{AUC} & \textbf{F1} \\
    \midrule
    5 (Early)                & 0.789  & 0.632 & 0.735 \\
    \textbf{12 (Middle)}     & \textbf{0.793}  & \textbf{0.657} & \textbf{0.743} \\
    20 (Late)                & 0.788  & 0.635 & 0.740 \\
    \bottomrule
  \end{tabular}
\end{table}

\vspace{3pt}
\noindent\textbf{SAE Feature Extraction.}  
For each selected layer $\ell \in \{5,12,20\}$, we apply SAELens to the residual‐stream activations and perform summation‐based pooling across token positions to obtain a pooled SAE vector $\mathbf{s}^{(\ell)} \in \mathbb{R}^{d^{(\ell)}}$, where $d^{(\ell)}$ is the activation dimensionality at layer $\ell$. We then apply feature selection using the tree-based method to get the top-$1500$ SAE features to train an $\ell_{2}$-regularized logistic regression classifier.

\vspace{3pt}
\noindent\textbf{Results and Analysis.}  
The results in Table \ref{tab:sae_layers} show that SAE activations from the intermediate layer (Layer 12) yield the strongest predictive performance for earnings surprise, with a weighted F1 of 0.743, an AUC of 0.657, and an accuracy of 0.793, which are all higher than the corresponding metrics from the early layer (Layer 5) and the late layer (layer 20). For the early layer, the features at this stage are still dominated by more surface‐level lexical and syntactic cues. Conversely, for the late layer, while high‐level abstractions are being formed, some of the fine‐grained financial signals critical for the earnings surprise task may become diffused deeper in the network.

These findings suggest that the \emph{middle‐layer features strike an optimal balance between surface‐level lexical–syntactic cues and deeper, high‐level semantic abstractions}. In practice, this means that extracting SAE activations from a middle layer can capture both the fine‐grained textual signals (e.g., sentiment‐laden phrases, domain‐specific terminology) and the broader contextual understanding (e.g., inter‐sentence relationships, narrative flow) necessary for accurate earnings surprise prediction.

\subsection{Interpretability Analysis}
To explore what our best models based on sparse autoencoder activations are capturing, we extracted the indices of the most important features selected for fitting linear classifiers in the two configurations: the 16K–width activations in layer 12 of Gemma 2-2B and the 131K–width activations in layer 20 of Gemma 2-9B. These indices were then queried in Neuronpedia~\citep{neuronpedia} to obtain human‐readable interpretation. Table \ref{tab:gemma2_2b_features} and \ref{tab:feature_activations_full} in the Appendix lists the top features for both models and their corresponding semantic annotations.

Inspection of these top features reveals a mix of abstract and concrete concepts. On one hand, there are features reflecting sentiment (e.g., positive evaluations or conditional phrasing), comparative constructs, and procedural language. On the other hand, a substantial portion of features pertains to geopolitics, finance and monetary values, scientific research metrics, medical and healthcare terminology, urban development and land‐economics lexicon, and legal standards. In the representative sentence shown in Table ~\ref{tab:gemma2_2b_features}, highlighted spans correspond to tokens that strongly activate the respective features, illustrating the interpretability and semantic co-occurrence of the learned representations. 

\section{Conclusions and Future Work}
In the paper, we propose SAE-FiRE, a novel framework that leverages SAEs to improve earnings surprise predictions from financial documents. By extracting key information while eliminating redundancy from lengthy financial texts, our approach effectively suppresses feature noise that might otherwise lead to model overfitting. Experimental results across three financial datasets demonstrate that SAE-FiRE  outperforms all baseline approaches across multiple performance metrics. Experiments also show that SAE-FiRE enhances model interpretability by identifying the specific semantic features. For future work, we plan to extend SAE-FiRE to multimodal settings by incorporating audio features from earnings calls alongside text and potentially adapt the framework to other financial prediction tasks.

\clearpage
\section*{Limitations}
Despite its strengths, SAE-FiRE has several limitations. First, our model relies on pretrained sparse autoencoders originally designed for general language tasks and not optimized for financial domain adaptation. Fine-tuning SAEs on financial corpora may yield further gains. Second, our feature selection methods, though effective, may discard weakly informative but complementary signals when enforcing strict sparsity. Third, our experimental setup focuses on binary classification, which simplifies the nuanced nature of earnings surprises that often vary in magnitude and impact. Finally, real-time deployment and inference efficiency remain open challenges due to the high dimensionality of activation vectors, especially in larger LLMs.

\bibliography{custom}

\clearpage
\appendix

\section{Related Work}





\noindent\textbf{Financial Prediction.}
Earnings calls provide rich forward-looking information, driving substantial research interest. Early studies, such as \citep{larcker2012detecting} and \citep{bushee2018linguistic}, focus on sentiment extraction and its impact on analyst forecasts. Recent advances apply deep learning to directly predict earnings surprises \citep{koval2023forecasting}. 
Multimodal models further enhance prediction accuracy by integrating text, audio, and network structures \citep{qin2019what, sawhney2020multimodal, sang2022predicting, yang2020beyond}. \citet{huang2023finbert} develop FinBERT, a domain-specific language model that significantly improves financial text understanding.

\vspace{3pt}
\noindent\textbf{Sparse Autoencoders.}
SAEs were originally developed for explainability purposes to understand the internal representations of LLMs by decomposing dense neural activations into interpretable sparse components \citep{shu2025surveysparseautoencodersinterpreting,gaoscaling}. \citet{ferrando2025knowledge} demonstrate that SAEs reveal meaningful latent directions in LLMs, enabling control over knowledge-awareness and hallucination behaviors. Similarly, \citet{he2025saif} introduce SAIF to steer instruction-following behaviors through interpretable latent features.

Although widely applied in NLP, SAE adoption in financial prediction remains limited. Given the noisy and high-dimensional nature of financial data, SAEs hold promise for improving robustness and predictive accuracy. In this paper, we extend SAEs to financial forecasting by integrating them into an earnings surprise prediction framework.

\section{Experiment Details}
\subsection{Datasets}\label{appendix:sec:datasets}
\textbf{Conference Call.} We manually collected English-language earnings call transcripts for the largest publicly traded U.S. firms from the Seeking Alpha website. The dataset consists of English-language transcripts from quarterly earnings conference calls of publicly traded U.S. companies.The transcripts were annotated with the
names of the speakers (executives and analysts) and content, offering rich, high-frequency textual and contextual data for studying corporate communication, sentiment, and strategic disclosures. To focus on liquid and economically significant firms, we restrict the sample to companies with market capitalizations exceeding \$1 billion and average daily trading volumes above \$50 million. Our sample included 9,324 earnings conference calls, between 2012 and 2014. The breakdown of our sample by year is presented in Table \ref{tab:summary_stats}. 

\noindent
\textbf{10\,-Q Report.} The second dataset comprises the \textit{Management Discussion and Analysis} (MDA) sections from firms’ quarterly (10Q) filings with the U.S. Securities and Exchange Commission (SEC). The MDA provides management’s narrative perspective on past performance, future outlook, and key business risks, offering rich forward-looking information. We focus on this section because it reflects direct managerial communication to shareholders. All texts are preprocessed to remove boilerplate language, tables, and footnotes to retain core narrative content.

\noindent
\textbf{FNSPID Nasdaq News.} The third dataset is \textit{FNSPID Financial News and Stock Price Integration Dataset}, a large-scale time-series dataset that integrates textual and market data \cite{dong2024fnspid}. It includes 29.7~million stock price observations and 15.7~million financial news records for 4,775 S\&P~500 firms from 1999 to 2023, collected from four major news websites. It provides summaries of financial news records generated using various methods as well. Given the scale of this corpus, processing full articles for a single firm over a quarter would result in inputs orders of magnitude longer than those in the other two datasets. Therefore, we use the dataset’s LSA summaries: for each firm–quarter we sort summaries chronologically and concatenate them into a single document formatted as a numbered list covering the entire quarter preceding the next EPS announcement. This retains broad topical coverage while keeping length manageable and comparable across datasets. FNSPID stands out for its scale, diversity, and sentiment signals derived from news, which have been shown to enhance transformer-based stock return predictions. It also provides a reproducible updating pipeline with code and documentation, making it a valuable resource for financial research.

\vspace{3pt}
\noindent\textbf{Financial Data.} We obtained Reported Earnings Per Share (EPS) and analyst consensus EPS forecasts from the IBES database. For each earnings call, we align realized EPS values with analyst consensus forecasts, allowing for the computation of earnings surprises. The combination of rich textual data with structured earnings performance metrics makes this dataset especially suitable for tasks such as managerial sentiment analysis and earnings surprise prediction.

\subsection{Evaluation Metric}
We evaluate classification performance using four standard metrics: Accuracy, Precision, Recall, and Weighted F1-score. These metrics are computed using a chronological train-validation-test split based on cutoff dates to avoid look-ahead. Additionally, we report the Area Under the Receiver Operating Characteristic Curve (AUC) to assess the model's ranking performance and sensitivity to classification thresholds.
To further assess model stability and feature effectiveness, we examine performance across various values of the selected top-k SAE features (500, 1000, 1500, 2000, 2500) for Gemma2-2B 16K SAE and (3000, 3500, 4000, 4500, 5000, 5500, 6000) for Gemma2-9B 131K SAE and Llama3.1-8B 131K SAE. All metrics are reported on a temporally held-out test set to prevent look-ahead bias.

\subsection{Baseline Details}\label{appendix-sec-baselines}
\noindent
\textbf{{Longformer.}}
    We fine‐tune the pretrained Longformer‐base‐4096 model \citep{beltagy2020longformer}. Each document is first split into chunks of up to 4096 tokens.  We encode each chunk independently with the pretrained `allenai/longformer‑base‑4096` to extract its position 0 token embedding (classification tasks with longformer commonly employs the beginning-of-sequence token) of the last hidden state, then aggregate the sequence of chunk embeddings with a bidirectional LSTM. We apply max‐pooling over the LSTM outputs to form a single document representation, and pass this through a two‑layer MLP classifier. The entire model is fine‑tuned end‑to‑end for 3 epochs.  

\noindent
\textbf{{Hierarchical FinBERT.}}
    We implement the Hierarchical FinBERT model from \citep{koval2023forecasting} for long-document  classification. The input earnings‑call transcript is split into segments of roughly 4,000 tokens. Each segment is independently encoded using FinBERT. The per‑segment embeddings are then aggregated through a  document‑level pooling layer to produce a full‑document representation, which is fed into a classification head to predict earnings surprises. 

\section{Training Details}
\subsection{Training Configuration}
All models were trained on NVIDIA H100 GPUs using CUDA 12.6 and Ubuntu 20.04. Model training was conducted in PyTorch 2.6 for MLP classifiers and \texttt{scikit-learn 1.6} for logistic regression and tree-based models. Each batch corresponds to one full transcript of up to 20,000 tokens, and random seeds were fixed to 42 for reproducibility. We used the AdamW optimizer with a learning rate of $1\times10^{-6}$, applied $\ell_2$ regularization, tuned via grid search, and incorporated dropout at rates of 0.5 for the 3-layer MLP; ReLU served as the activation function, and training employed early stopping with a patience of 20 epochs based on validation AUC or loss.

\subsection{Training Process}

Each input text is passed through the pre-trained LLM equipped with SAELens to extract residual-stream activations. We apply a token-wise sum-pooling operation across all positions to obtain a fixed-length sparse vector $s \in \mathbb{R}^d$ per document. 

Next, we apply univariate feature selection using three statistical criteria: ANOVA F-test and Tree-based methods. From these scores, we retain the top-$k$ features where $k \in \{500, 1000, 1500, 2000, 2500\}$ when using Gemma 2-2B 16K SAE, and $k \in \{3000, 3500, 4000, 4500, 5000, 5500, 6000\}$ when using Gemma 2-9B 131K SAE and when using Llama3.1-8B 131K SAE. The resulting feature vector is then passed to a logistic regression or MLP classifier.

Models are trained using a chronological split into training, validation, and test sets based on cutoff dates, ensuring the training data never includes future information. Hyperparameters are tuned on the validation set, and final results are reported on the held-out test set.

\begin{figure*}[htb]
  \centering
  \includegraphics[width=\textwidth]{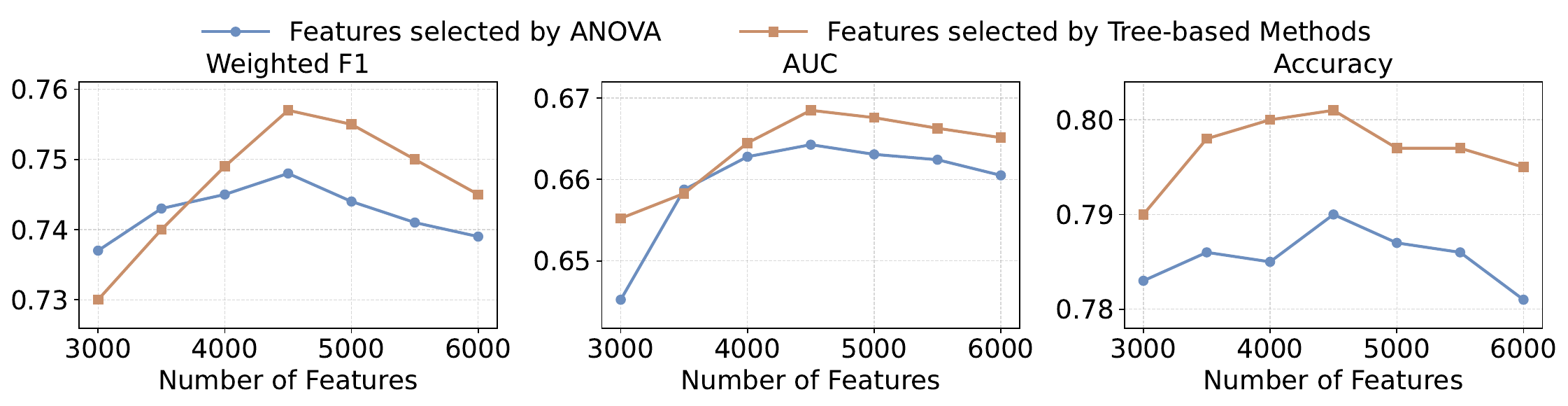}
  \caption{
    Weighted F1, AUC, and Accuracy at different numbers of selected features. (Gemmma2-9B 131K)
  }
  \label{fig:num-feature-comparison1}
\end{figure*}

\begin{figure*}[htb]
  \centering
  \includegraphics[width=\textwidth]{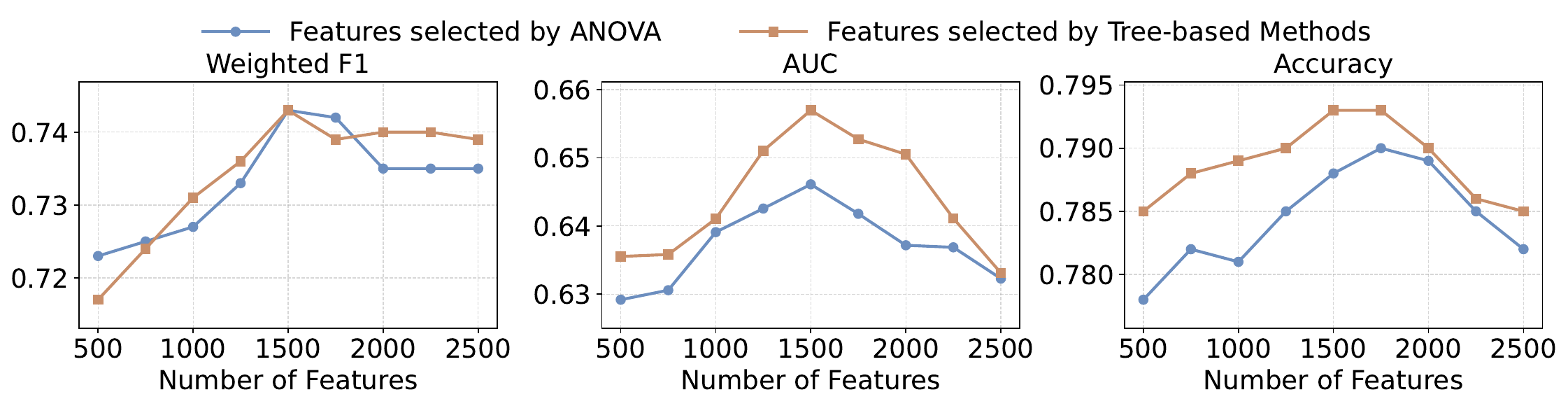}
  \caption{Weighted F1, AUC, and Accuracy at different numbers of selected features (Gemma2-2B 16K)}
  \label{fig:num-feature-comparison3}
\end{figure*}

\begin{figure}[htb]
  \centering
  \includegraphics[width=\columnwidth]{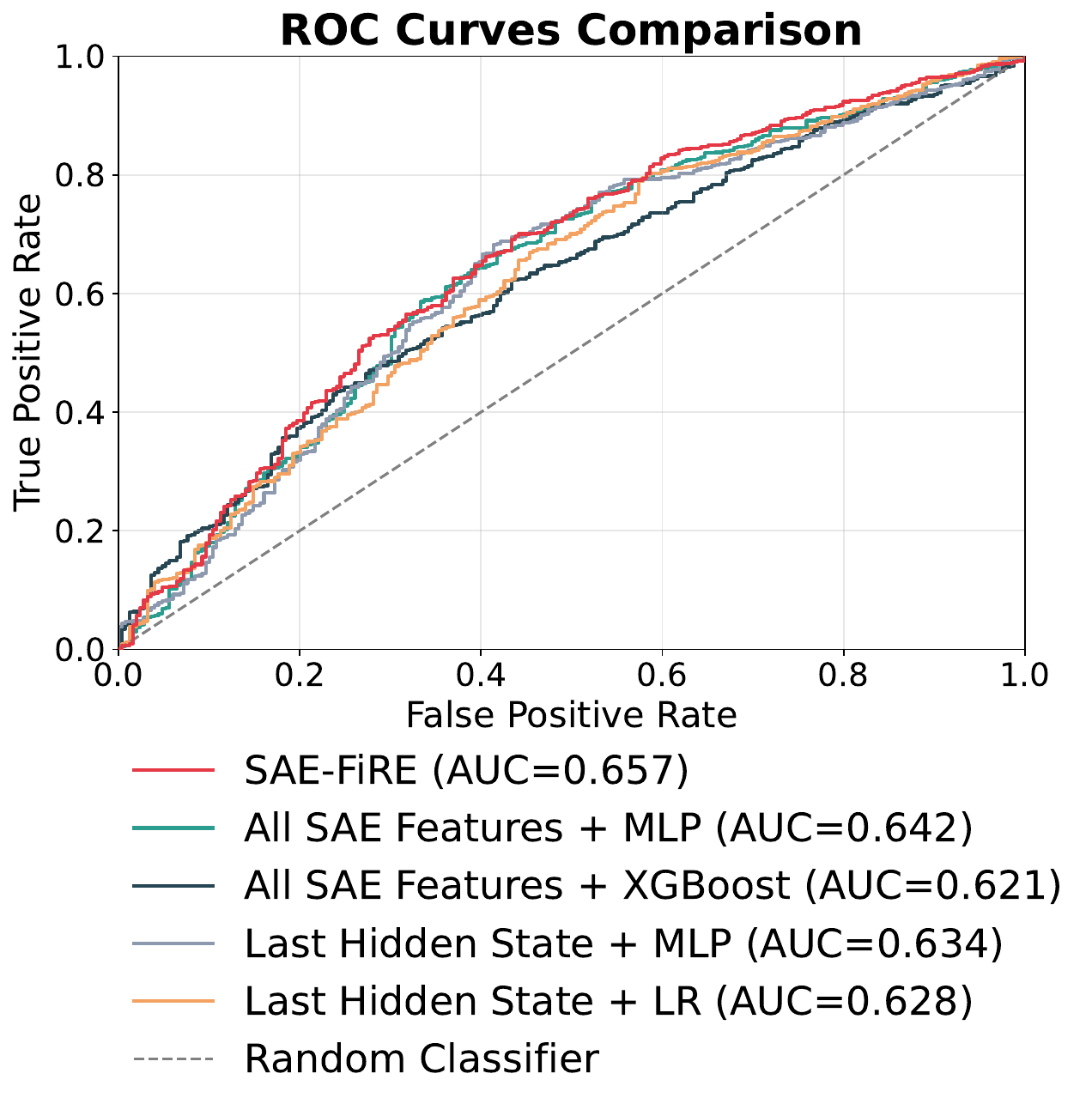}
  \caption{ROC Curves of our model and baseline models}
  \label{fig:model-baseline-comparison1}
\end{figure}

\begin{figure}[htb]
  \centering
  \includegraphics[width=\columnwidth]{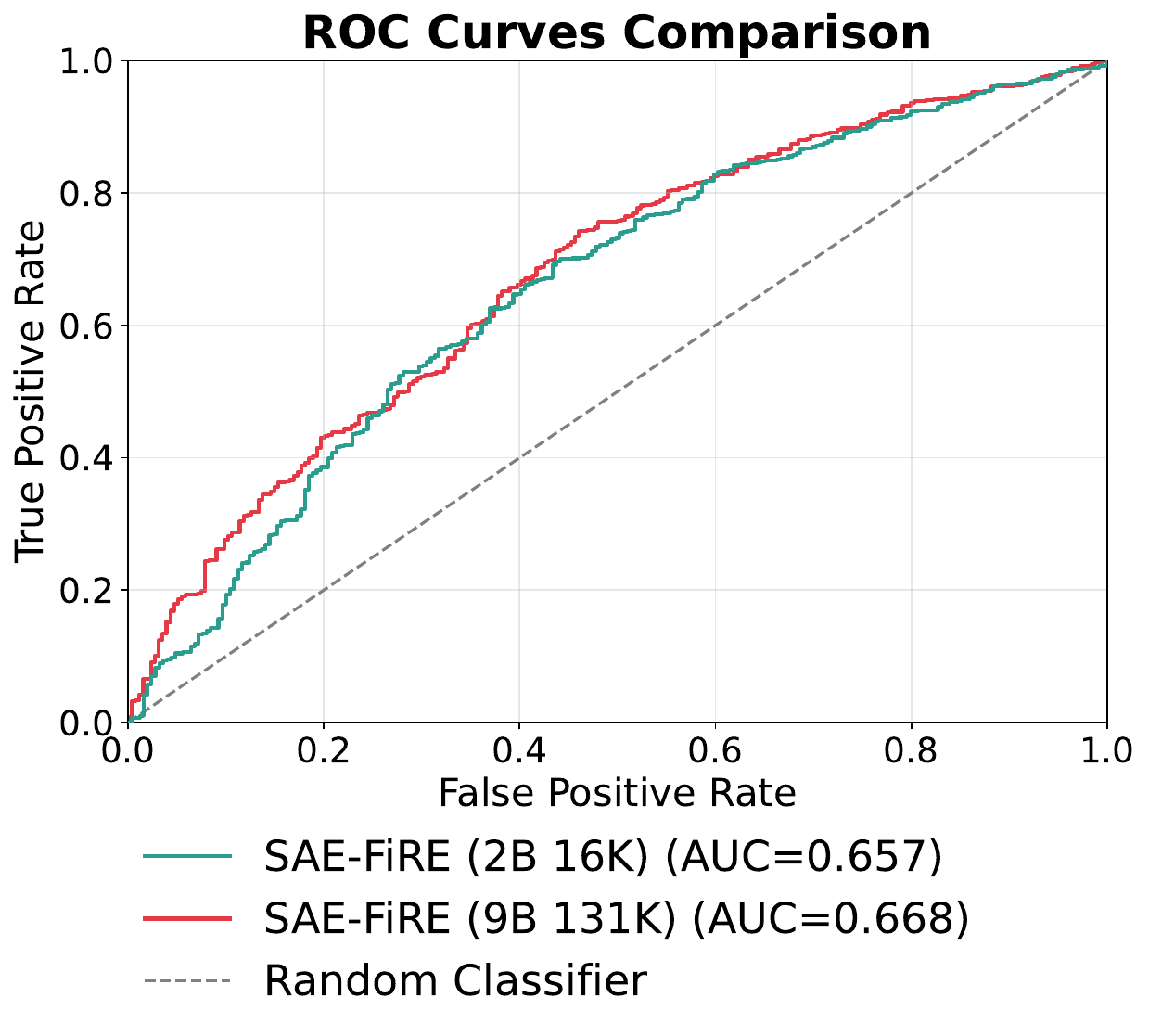}
  \caption{ROC Curves of our model with different large language model sizes}
  \label{fig:model-size-comparison1}
\end{figure}

\begin{figure}[htb]
  \centering
  \includegraphics[width=\columnwidth]{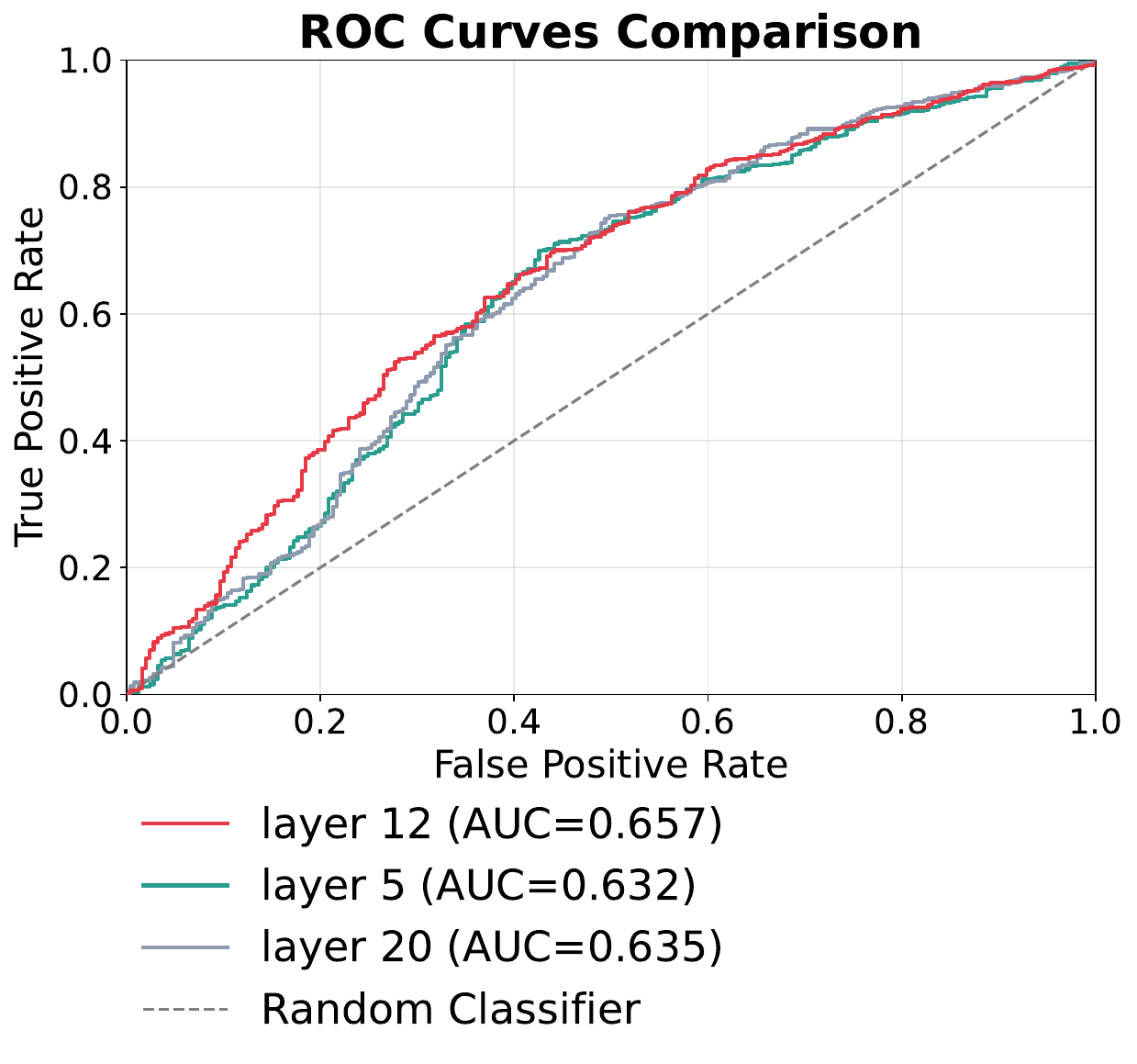}
  \caption{ROC Curves of our model with SAE activations from residual streams of different layers}
  \label{fig:model-layer-comparison1}
\end{figure}

\begin{table*}[ht]
  \centering   
  \begingroup
  \renewcommand{\arraystretch}{1.2}
    {\fontsize{10 pt}{12pt}\selectfont
      \begin{tabular}{@{} l l p{12cm} @{}}
        \toprule
        \textbf{Model} & \textbf{Feature} & \textbf{Interpretation (Neuronpedia)} \\
        \midrule
        \multirow{5}{*}{Gemma 2-2B 16K}
          & 4334   & legal terms and phrases related to contracts and warranties \\
          & 3820   & financial terms and references to wealth or monetary values \\
          & 1608   & phrases related to affirmation, recommendations, and positive evaluations \\
          & 4595   & phrases related to customer satisfaction and support \\
          & 4961   & references to oil and gas exploration in a geographical context \\
          & 5096   & references to urban environments and their characteristics \\
          & 6751   & terms related to political processes and events \\
          & 2551   & technical terms and phrases related to data transfer and processing \\
          & 445   & phrases related to procedural instructions and configurations \\
          & 7334   & terms related to the development and improvement of mechanisms or products \\
          & 2401   & references to awards and achievements \\
          & 3127   & information related to medical interventions and their effectiveness or outcomes \\
          & 3182   & aspects related to quality, efficacy, and performance metrics in various contexts \\
          & 5954   & phrases indicating a search or request for information that has not been successfully fulfilled \\
          & 6873   & terms related to experimental procedures and comparisons across different groups or times \\
        \addlinespace
        \cmidrule(lr){1-3}
        \multirow{5}{*}{Gemma 2-9B 131K}
          & 114574 & terms related to real estate and urban development \\
          & 36381  & topics related to regulations and their impact on the economy \\
          & 100994 & monetary values and project costs associated with projects \\
          & 54667  & specific numerical values or measurements in a scientific context \\
          & 65854  & specific medical conditions and related terminologies \\
          & 81238  & geographical locations and their corresponding identifiers \\
          & 118806  & positive descriptions of experiences or situations \\
          & 129720  & conditional phrases and expressions of uncertainty \\
          & 110418  & terms related to transportation logistics and procedures \\
          & 24250  & concepts related to economic supply dynamics \\
          & 54667  & specific numerical values or measurements in a scientific context \\
          & 103050  & terms related to inventions and innovations in the United States \\
          & 1342  & keywords related to statistical methods and data handling \\
          & 16491  & phrases indicating factual information or definitive assertions \\
          & 92322  & terms and concepts related to legal standards and calculations \\
        \bottomrule
      \end{tabular}
    }
  \endgroup
  \caption{Top SAE Activation Features and Their Interpretations}
  \label{tab:feature_activations_full}
\end{table*}





\section{More Experimental Results}

\subsection{Ablation Studies Based on Gemma2-2B}
We present ablation studies of SAE-FiRE within the Gemma2-2B (16K) setting, isolating the effects of sum-pooled SAE representations and tree-based feature selection on discriminative performance. Figure~\ref{fig:model-baseline-comparison1} presents the ROC curves and corresponding AUC scores for each method, providing insight into their true positive rates under varying false positive thresholds. Our proposed SAE-FiRE model attains the highest AUC of 0.657, outperforming all baselines across the ROC curve. This result confirms the benefit of combining sum-pooled SAE representations with tree-based feature selection, which enhances class separability and reduces noise from irrelevant dimensions. Models trained on the full 16K SAE vectors without feature selection perform worse than SAE-FiRE. The MLPclassifier achieves an AUC of 0.642, while the XGBoost-based classifier trails further at 0.621. This decline when using all SAE features without selection can be attributed to overfitting and the inclusion of noisy or redundant dimensions. Probing the final hidden state with MLP or logistic regression results in AUC scores of 0.634 and 0.628, respectively. While these approaches outperform XGBoost on raw SAE features, they still fall short of SAE-FiRE, suggesting that hidden state vectors lack the global structural information encoded in pooled SAE activations.

\subsection{Cross-Model Comparison of Feature Selection Strategies}
We conduct a cross-model comparison to evaluate how feature selection method and dimensionality affect performance across two SAE models, Gemma 2-9B 131K and Gemma 2-2B 16K. Results are reported in Figure~\ref{fig:num-feature-comparison1} and Figure ~\ref{fig:num-feature-comparison3}, which presents trends in weighted F1, AUC, and accuracy under varying numbers of selected features. Across both models and all three evaluation metrics, tree-based feature selection methods demonstrate superior performance compared to ANOVA. This advantage is particularly pronounced in AUC and accuracy, where tree-based selectors capture higher-order feature interactions and class-discriminative signals more effectively. In contrast, AUC and accuracy are more stable and favor tree-based selection across the board, especially at smaller feature sizes, where interaction modeling provides greater value. These results underscore two key takeaways: (1) tree-based feature selection is generally more effective across models and metrics, and (2) the optimal number of selected SAE features must be tailored to the base model’s capacity and dimensionality. Smaller models benefit from more aggressive feature pruning, while larger models require broader coverage to fully leverage their latent representations.


\subsection{Impact of Language Model Size on Classification Performance}
To assess how the size of the underlying language model influences classification quality, we compare the ROC curves of our SAE-FiRE framework using Gemma 2-2B (16K dimensions) and Gemma 2-9B (131K dimensions). The results are presented in Figure~\ref{fig:model-size-comparison1}. As shown in the ROC curves, the model based on Gemma 2-9B consistently maintains a higher true positive rate across most false positive thresholds compared to Gemma 2-2B. The AUC achieved by the larger model is 0.668, slightly outperforming the 2-2B variant, which attains an AUC of 0.657. This suggests that increased model capacity translates to better latent feature representations that enhance downstream predictive performance. 

\subsection{Effect of SAE Activation Layer on Classification Performance}
To investigate the impact of layer selection on the quality of SAE features, we extract activations from different residual streams (layers 5, 12, and 20) of the transformer and compare their ROC curves. Figure~\ref{fig:model-layer-comparison1} reports the classification performance of our model using SAE-FiRE features from each layer. Among the three layers evaluated, SAE features extracted from layer 12 result in the highest AUC of 0.657. The corresponding ROC curve consistently dominates those of layers 5 and 20, especially in the low false positive rate regime. This indicates that intermediate layers may offer a more balanced trade-off between semantic abstraction and retention of fine-grained input information. SAE-FiRE using layer 5 activations achieves a lower AUC of 0.632, suggesting that earlier layers may not capture sufficient semantic context for reliable classification. Similarly, activations from the deeper layer 20 yield an AUC of 0.635, which also falls short of layer 12. This performance drop may be attributed to over-specialization or noise accumulation in deeper layers.

\section{Case Analysis}
\label{sec:case_study}

\subsection{Structure of Earnings Conference Call}
An earnings conference call typically contains four sections: an introduction from the operator and the executive host, an opening section from the executive presenter, Q\&A between analysts and the executive presenter, and some closing address from the executive host and the operator. 

\subsubsection{Introduction Section}
\begin{figure}[htb]
  \centering
  \includegraphics[width=\columnwidth]{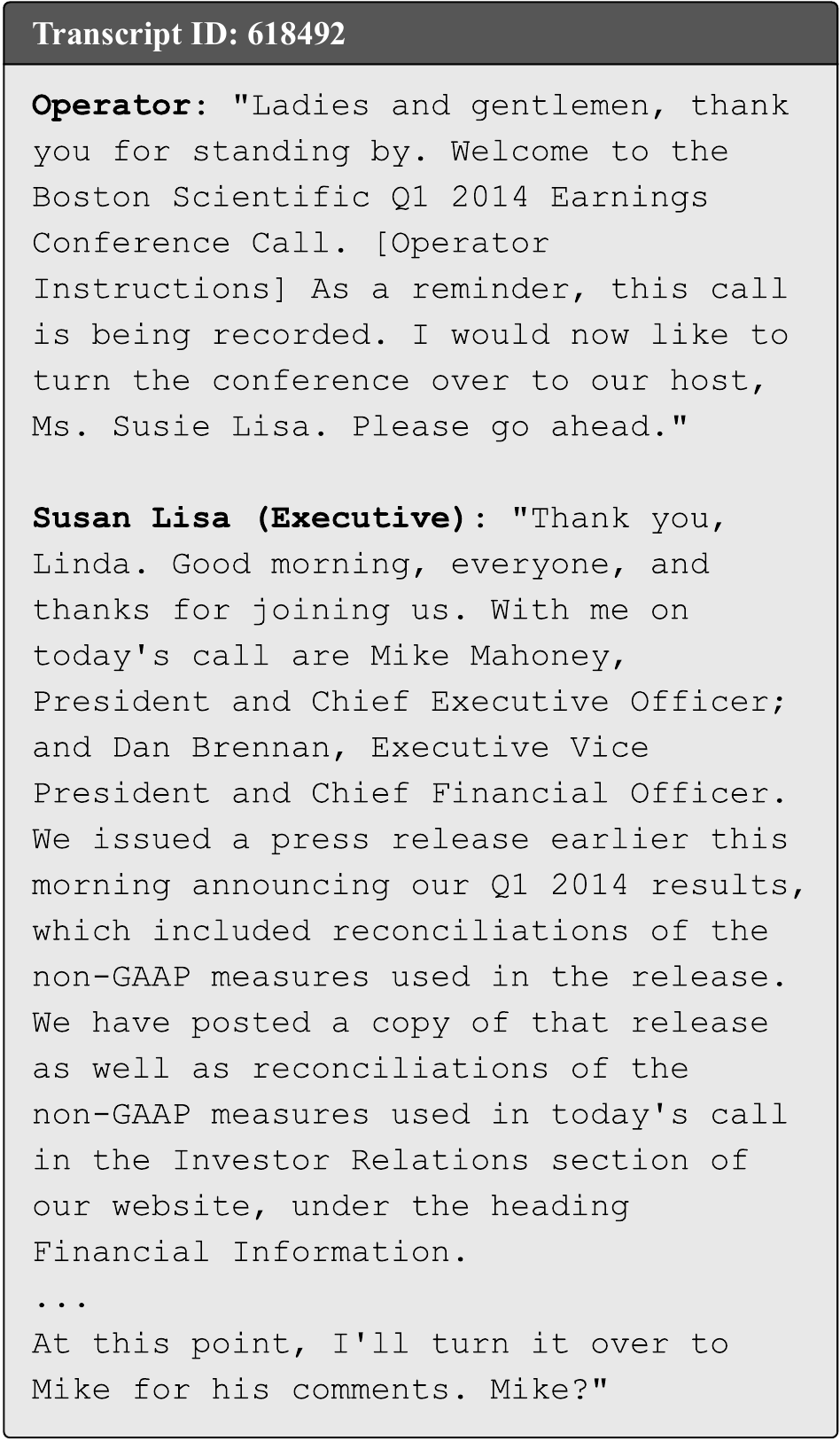}
  \caption{Transcript ID 618492 - Introduction}
  \label{fig:618492_introduction_text}
\end{figure}

The introduction opens with the operator’s welcome and call logistics, then the executive host thanks participants, introduces the leadership team, references the posted earnings release, and hands the presentation over to the CEO. Figure \ref{fig:618492_introduction_text} shows the introduction section for transcript 618492.

\subsubsection{Opening Section}
\begin{figure}[htb]
  \centering
  \includegraphics[width=\columnwidth]{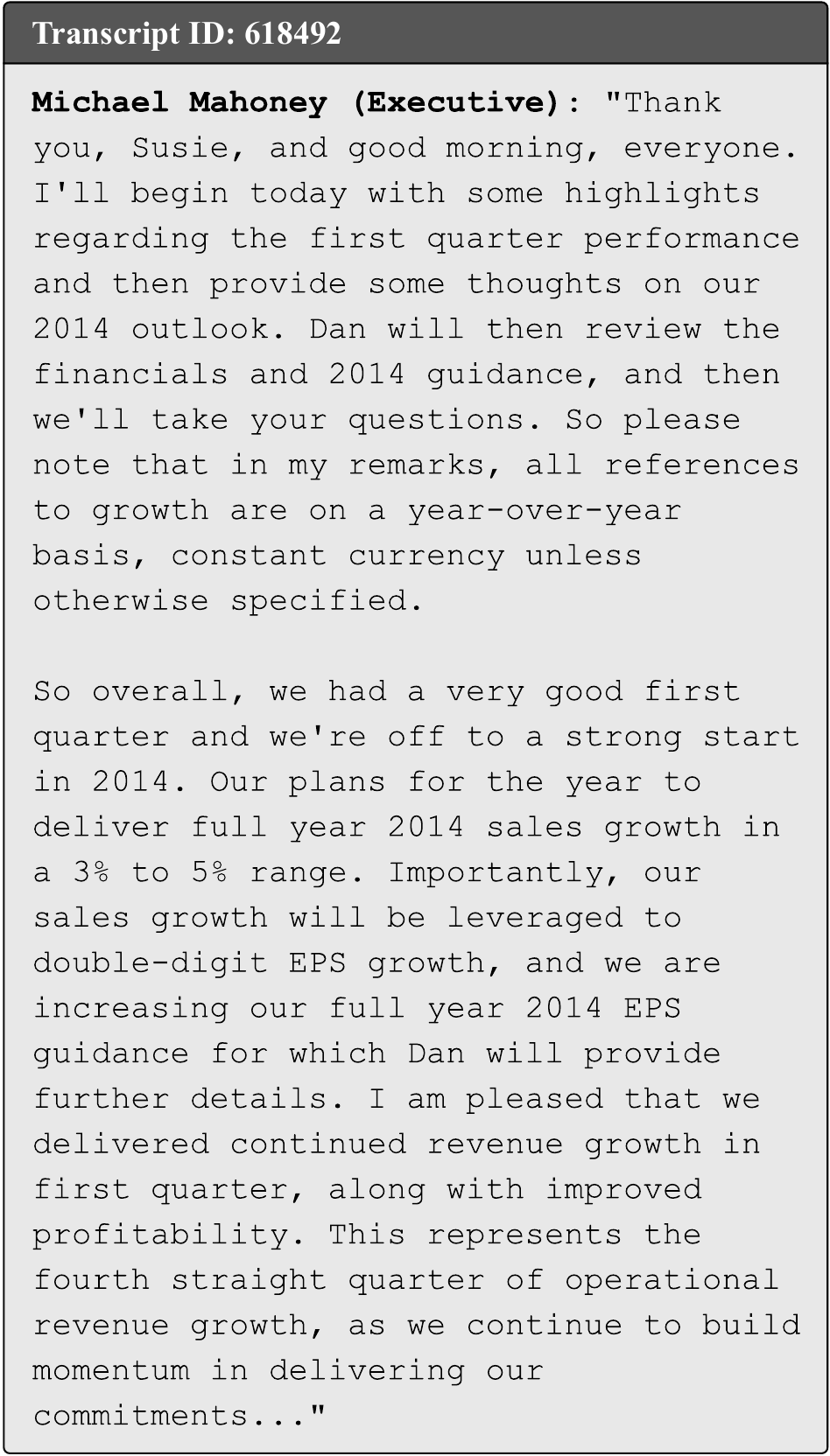}
  \caption{Transcript ID 618492 - Opening}
  \label{fig:618492_opening_text}
\end{figure}

The opening typically begins with a greeting and agenda overview, clarifies reference conventions, then highlights quarterly performance and full-year guidance before transitioning to the detailed financial review and Q\&A. Figure \ref{fig:618492_opening_text} shows the opening section for transcript 618492.

\subsubsection{Q\&A Section}
The Q\&A section comprises the bulk of the earnings conference transcript and includes multiple rounds of questions and answers. Each round begins with the operator introducing the analyst, the analyst posing a targeted question, and the executive providing a detailed response that addresses strategic considerations and market context. Figure \ref{fig:618492_qna_text} shows two questions and answers found in transcript 618492.

\subsubsection{Closing Section}
\begin{figure}[htb]
  \centering
  \includegraphics[width=\columnwidth]{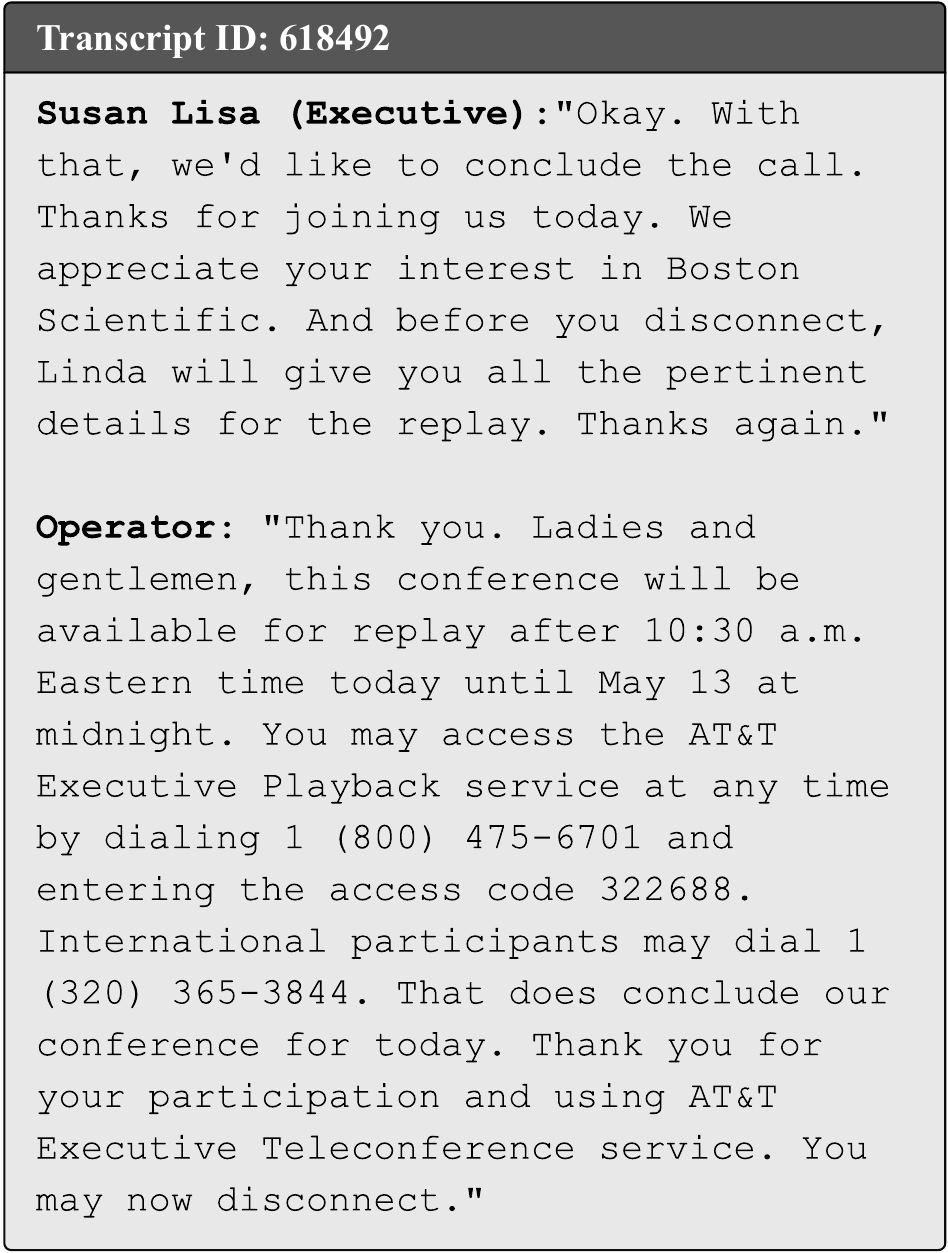}
  \caption{Transcript ID 618492 - Closing}
  \label{fig:618492_closing_text}
\end{figure}

As shown in Figure~\ref{fig:618492_closing_text}, an earnings conference call transcript usually ends with the executive thanking participants and introducing the replay details, after which the operator provides the replay access instructions, availability window, dial-in numbers, and a formal closing before disconnection.

\subsection{Activation of SAE Features in Transcripts}

To interpret which features most influenced classification decisions, we examined the top-ranked SAE activation indices selected by our feature selection pipeline. These indices were mapped to human-readable semantics using \texttt{Neuronpedia} \citep{neuronpedia}. We then use examples to show the activation of chosen features on transcripts and as well as the interpretations given by Neuronpedia.

\subsubsection{Top SAE Features for Earning Surprise Classification}

Table~\ref{tab:feature_activations_full} lists the top features identified by our ranking method as most critical for classification. The top features extracted from the Gemma2-9B SAE with 131K dimensions included activations associated with legal terminology, financial metrics, conditional phrases, and domain-specific jargon. This suggests that our SAE-FiRE framework captures both abstract semantic cues such as tone and modality, and concrete financial indicators relevant to earnings surprise forecasting. These features not only improve performance but also enhance interpretability by surfacing latent decision factors. 

\subsubsection{The sparsity of SAE Activation and Noises}
\begin{figure*}[htb]
  \centering
  \includegraphics[width=\textwidth]{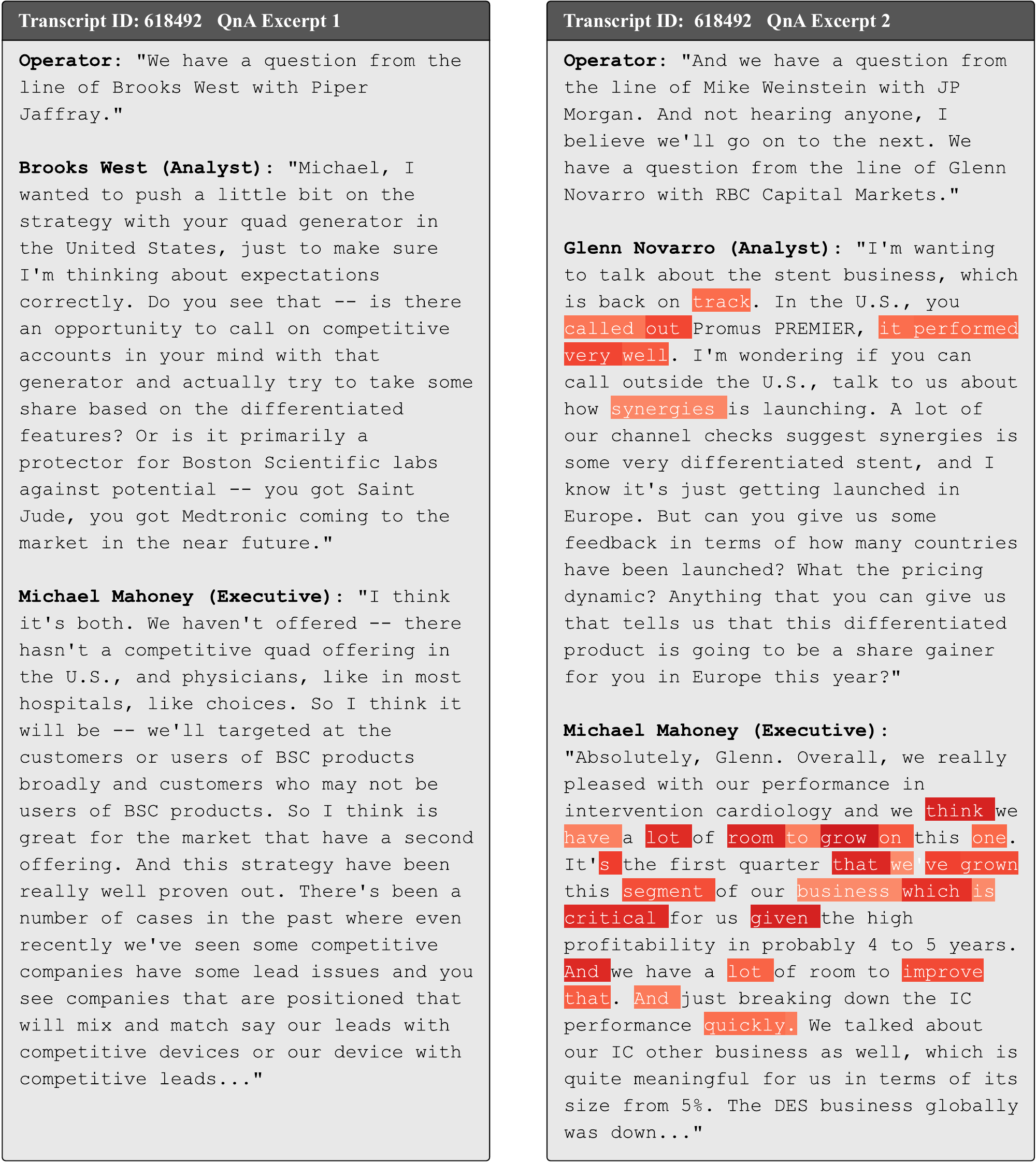}
  \caption{Transcript ID 618492 - Two excerpts of Q\&A and Comparison of Firing of SAE Feature Index 16491: phrases indicating factual information or definitive assertions}
  \label{fig:618492_qna_text}
\end{figure*}

The activation of individual SAE features which are the signals the model uses to classify positive or negative surprise can be highly context-specific and targeted. A given feature may fire only in specific sections of the text and be entirely absent elsewhere. For example, in the two Q\&A excerpts shown in Figure~\ref{fig:618492_qna_text}, feature 16491 (phrases indicating factual information or definitive assertions) is completely absent in the first question–answer pair but activates frequently in the second, despite both coming from the same transcript. By contrast, noise can crop up throughout different parts of the text, adding to the difficulty for classification. In this regard, those focused SAE activations carry more information and truly serve as reliable indicators of positive versus negative surprise.

\subsubsection{Important SAE Features For Positive and Negative Surprises}

\begin{figure*}[htb]
  \centering
  \includegraphics[width=\textwidth]{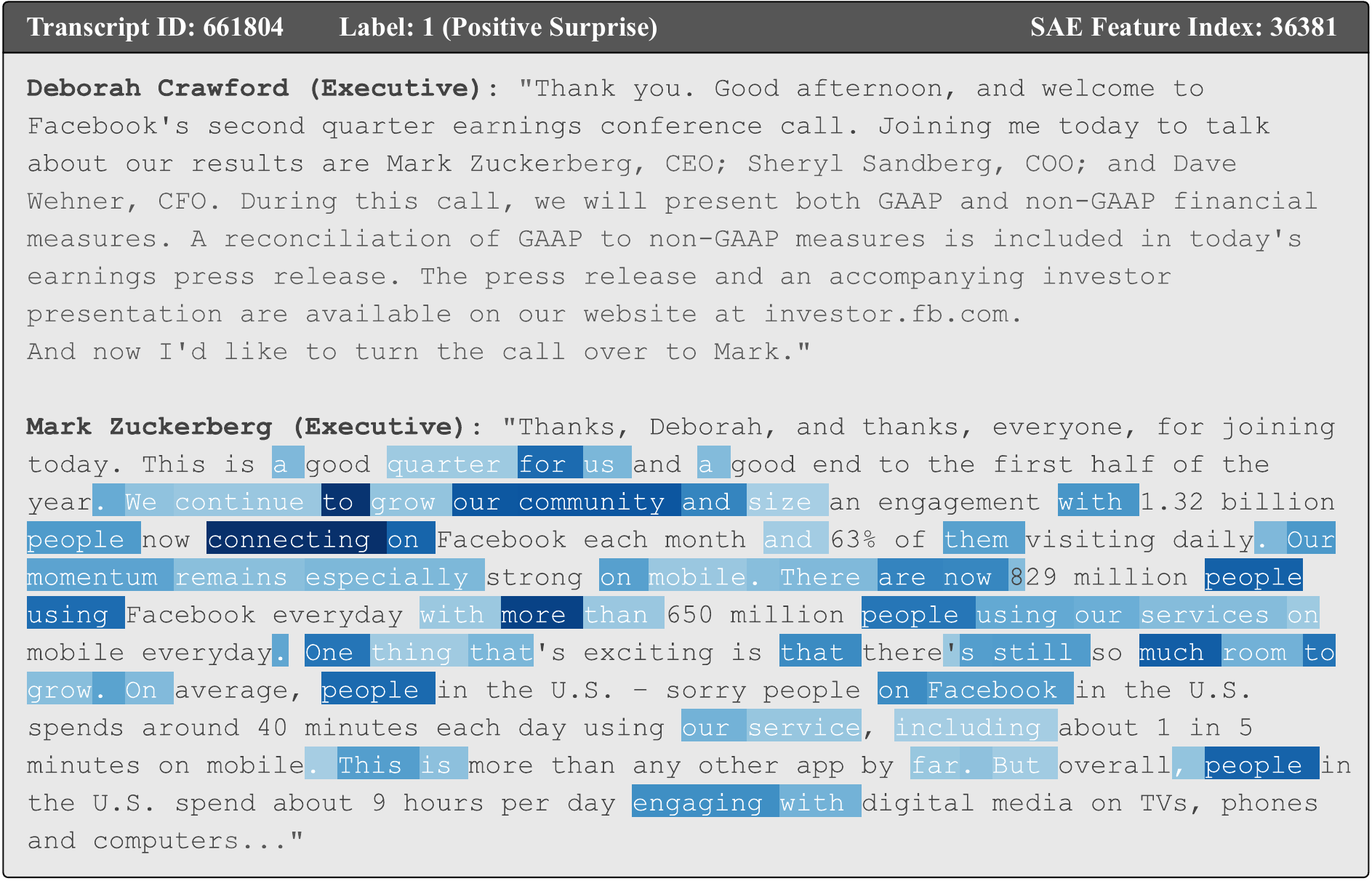}
  \caption{Transcript ID 661804's Activation on Feature Index 36381: topics related to regulations and their impact on the economy}
  \label{fig:36381_661804_text}
\end{figure*}

\begin{figure*}[htb]
  \centering
  \includegraphics[width=\textwidth]{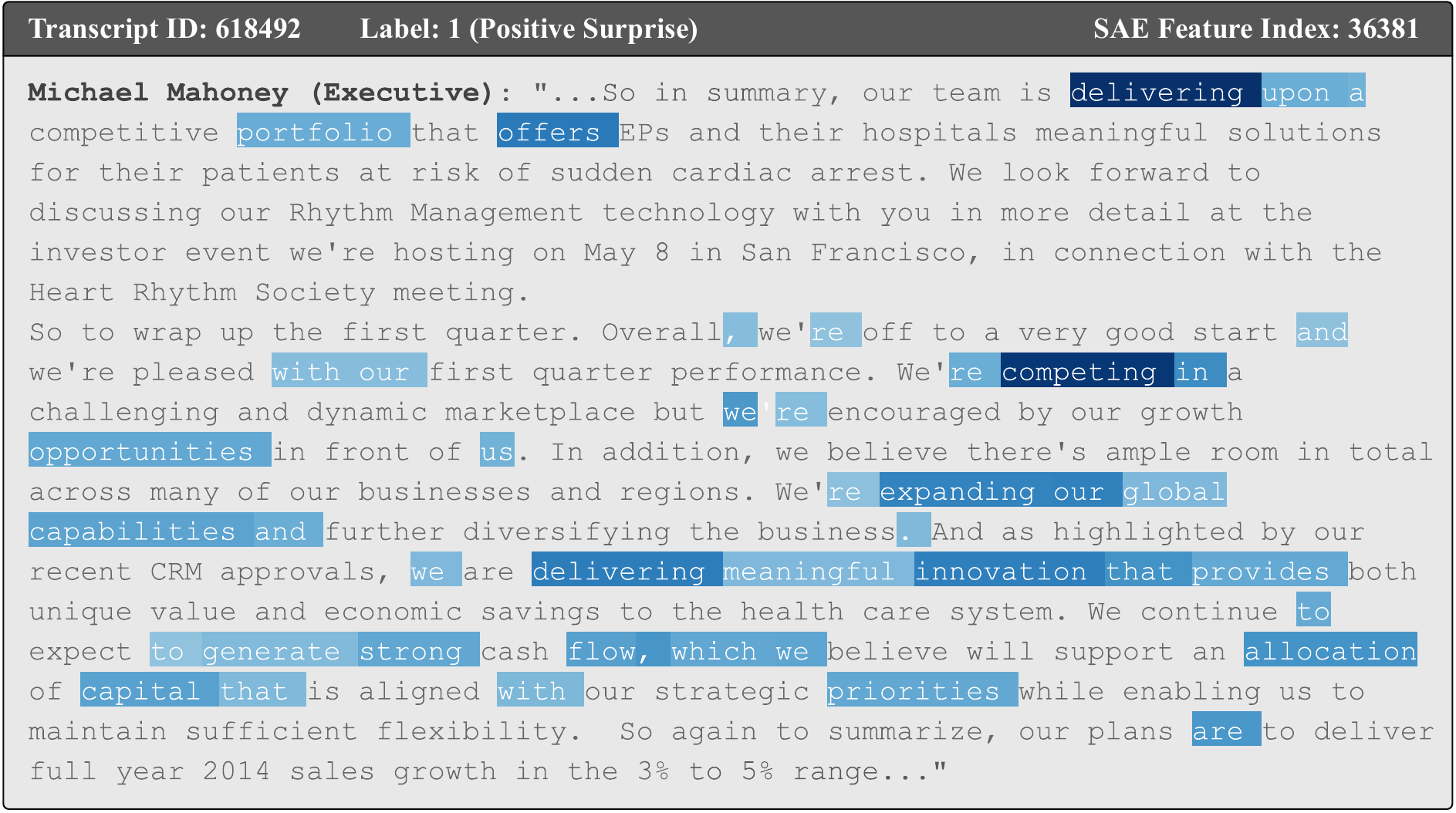}
  \caption{Transcript ID 618492's Activation on Feature Index 36381: topics related to regulations and their impact on the economy}
  \label{fig:36381_618492_text}
\end{figure*}

\begin{figure*}[htb]
  \centering
  \includegraphics[width=\textwidth]{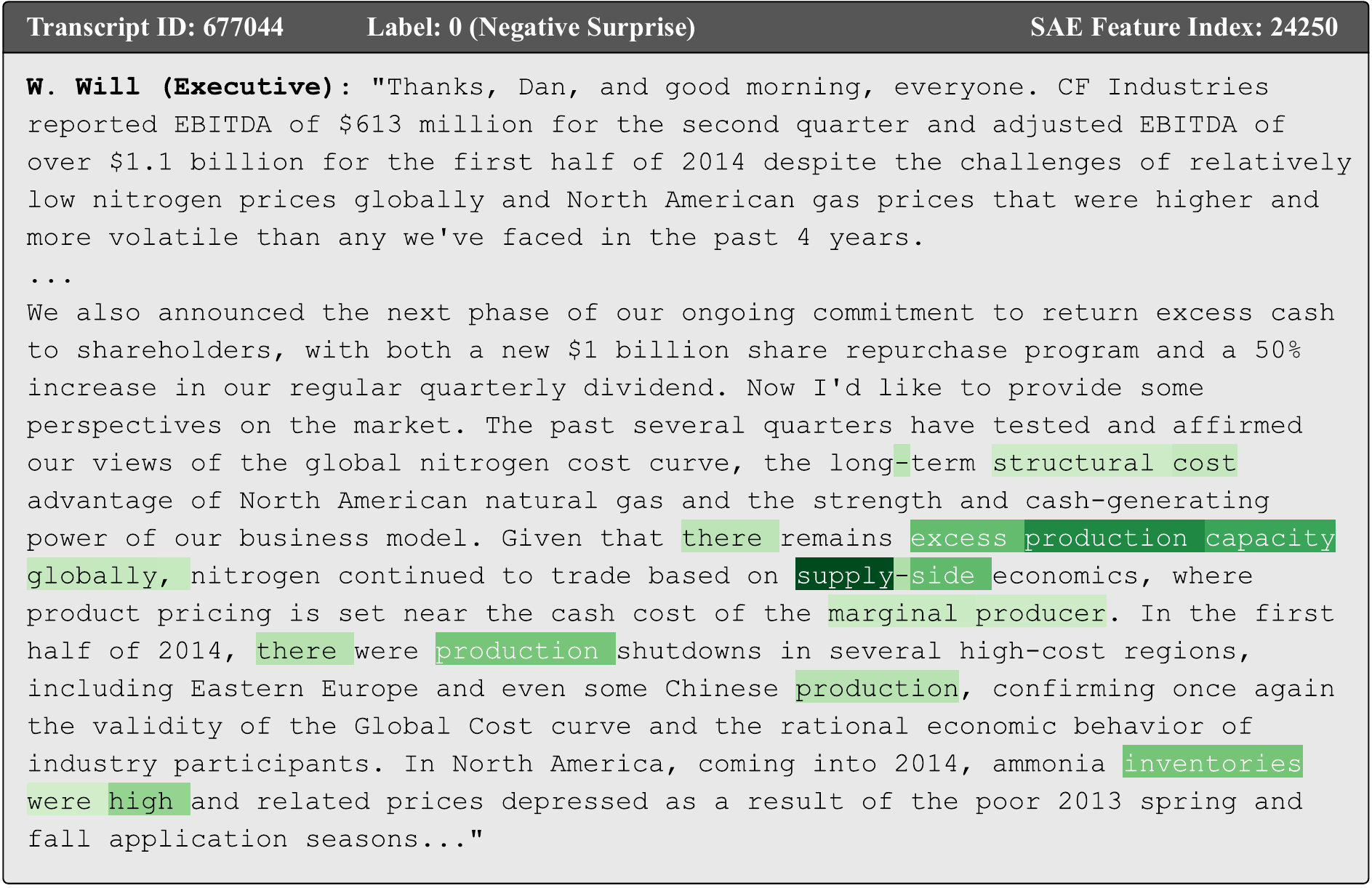}
  \caption{Transcript ID 677044's Activation on Feature Index 24250: concepts related to economic supply dynamics}
  \label{fig:24250_677044_text}
\end{figure*}

\begin{figure*}[htb]
  \centering
  \includegraphics[width=\textwidth]{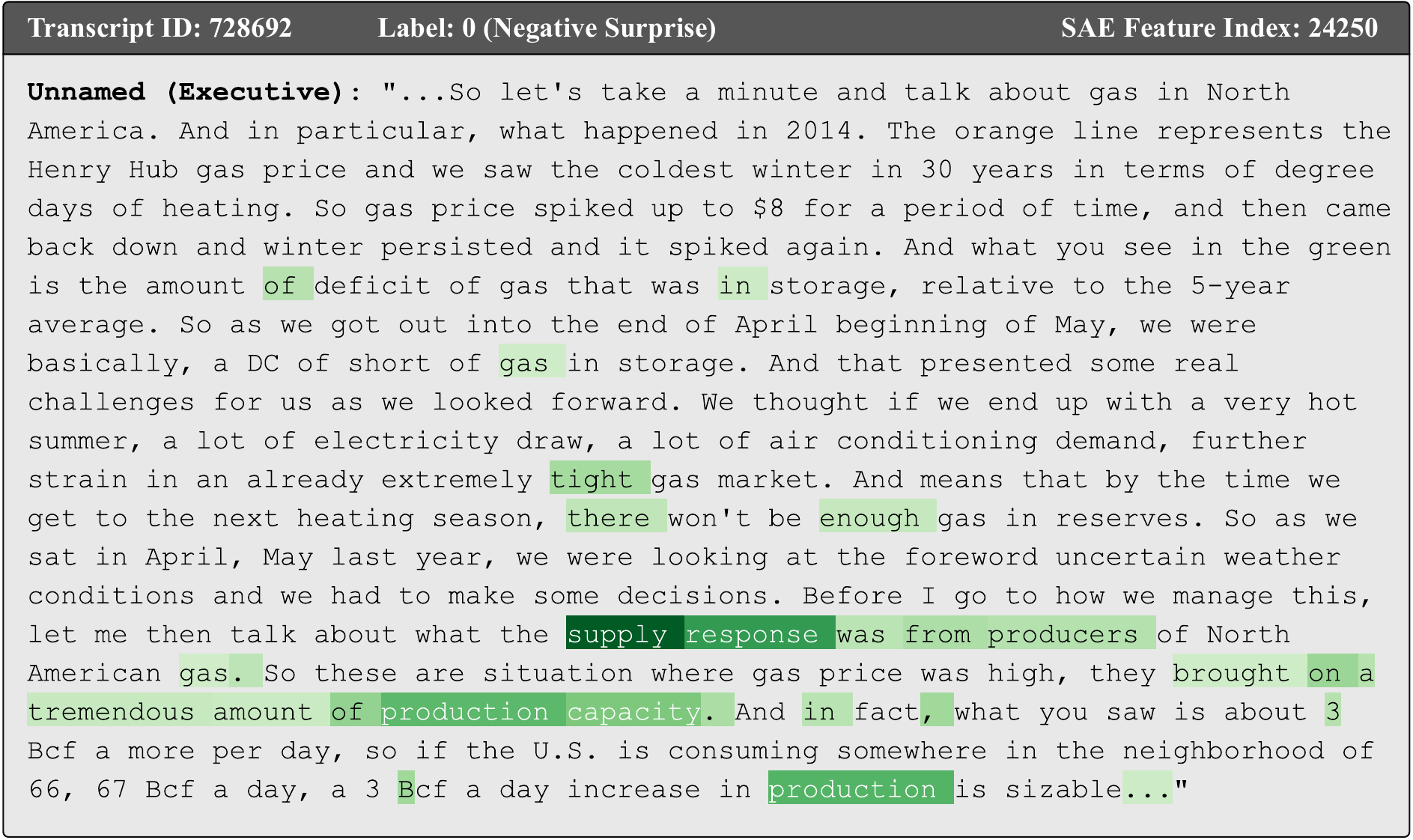}
  \caption{Transcript ID 728692's Activation on Feature Index 24250: concepts related to economic supply dynamics}
  \label{fig:24250_728692_text}
\end{figure*}

In Figure~\ref{fig:36381_661804_text} and Figure~\ref{fig:36381_618492_text}, we illustrate how feature 36381 (topics related to regulations and their impact on the economy) consistently activates in contexts the model associates with positive surprise. Meanwhile, Figure~\ref{fig:24250_677044_text} and Figure~\ref{fig:24250_728692_text} show instances where feature 24250 (concepts related to economic supply dynamics) activates in passages of transcripts the model interprets as negative surprises.

\section{Future Work}
\label{sec:future_work}


We plan to extend SAE-FiRE in several directions. First, we will incorporate multimodal inputs, including audio features from the earnings calls, to better capture vocal tone, hesitations, and emphasis, important signals in managerial communication. Second, we aim to adapt the framework to cross-lingual corporate disclosures, such as earnings calls from non-U.S. firms, by leveraging multilingual sparse autoencoders. Finally, we will explore applying SAE-FiRE to related financial forecasting tasks, e.g., credit risk prediction, guidance revision detection, or stock return volatility modeling.

\end{document}